\documentclass[fleqn,usenatbib]{mnras}
\usepackage[T1]{fontenc}

\DeclareRobustCommand{\VAN}[3]{#2}
\let\VANthebibliography\thebibliography
\def\thebibliography{\DeclareRobustCommand{\VAN}[3]{##3}\VANthebibliography}

\usepackage{graphicx}	% Including figure files
\usepackage{amsmath}	% Advanced maths commands
\usepackage{amssymb}	% Extra maths symbols
\usepackage{subcaption} % subfigures

\usepackage{newtxtext,newtxmath}
\title[The Extended Emission of QSO E1821+643]{The Kinematics and Ionization Structure of the Extended Emission Line Region of QSO E1821+643}

\author[S. A. Rosborough et al.]{
Sara A. Rosborough,$^{1}$\thanks{E-mail: sarosboro@gmail.com}
A. Robinson,$^{1}$
and T. Seelig$^{2}$
\\
$^{1}$School of Physics \& Astronomy, and Laboratory for Multiwavelength Astrophysics, Rochester Institute of Technology, Rochester, NY, USA\\
$^{2}$National Radio Astronomy Observatory, Socorro, NM, USA\\
}

\date{Accepted 2022 July. Received 2022 June; in original form 2021 September 10}

\pubyear{2021}

\begin{document}
\label{firstpage}
\pagerange{\pageref{firstpage}--\pageref{lastpage}}
\maketitle

% limit 250 words
\begin{abstract}
The most luminous quasars are created by major, gas-rich mergers and E1821+643, an optically luminous quasar situated at the center of a cool-core cluster, appears to be in the late stages of the post-merger blowout phase. This quasar is also identified as a gravitational recoil candidate, in which the supermassive black hole (SMBH) has received a recoil kick due to anisotropic emission of gravitational waves during the coalescence of a progenitor SMBH binary.  We analyze long-slit spectra of the extended, ionized gas surrounding E1821+643 to study its kinematics and ionization.  We have identified three kinematically distinct components, which we associate, respectively, with a wide-angle polar wind from the nucleus, kinematically undisturbed gas, and a redshifted arc-like structure of gas, at a distance of 3-4\arcsec~(13-18 kpc) from the nucleus.  The latter component coincides with the northern and eastern extremities of an arc of [OIII] emission seen in HST images. This feature could trace a tidal tail originating from a merger with a gas-rich galaxy to the South-East of the nucleus, whose presence has been inferred by Aravena et al. from the detection of CO emission.  Alternatively, the arc could be the remnant of a shell of gas swept-up by a powerful quasar wind.  The emission line ratios of the extended gas are consistent with photoionization by the quasar, but a contribution from radiative shocks cannot be excluded.    
\end{abstract}


%%%%%%%%%%%%%%%%%%%%%%%%%%%%%%%%%%%%%%%%%%%%%%%%%%

%%%%%%%%%%%%%%%%% BODY OF PAPER %%%%%%%%%%%%%%%%%%

\section{Introduction}
\label{sec:intro}
%This is a simple template for authors to write new MNRAS papers.
%See \texttt{mnras\_sample.tex} for a more complex example, and \texttt{mnras\_guide.tex} for a full user guide.

The luminous quasar, E1821+643 (henceforth, E1821), has a number of highly unusual characteristics that make it an object of great interest. It appears to be a post-merger system, which is transitioning from an ultra-luminous infrared galaxy to an optically luminous quasar \citep{Aravena2011} and its host galaxy resides at the center of a rich cool-core cluster (CCC) \citep{Russell2010}.  E1821 has also been identified as a potential supermassive black hole (SMBH) binary \citep{Blundell2001TheStructure}, which may have already merged, imparting a gravitational recoil kick to the product SMBH \citep{Robinson2010}. 

The most luminous quasars are associated with major, gas-rich mergers and the evolution process is comprehensively described by \citet{Hopkins2008AActivity}.  During the coalescence phase, gas inflows to the center and the luminosity of the merging system  is dominated by starbursts and an (x-ray) active galactic nucleus (AGN) is buried in the gas.  The gas feeds the rapidly growing, newly merged, central SMBH (or binary SMBH).  Briefly, the AGN light dominates in the "blowout" phase as radiation pressure driven winds expel the remaining gas and dust.  With the dust removed, a luminous quasar is revealed.  Eventually, the quasar's luminosity fades, as do the tidal tails and evidence of the merger, finally leaving a dust-free, elliptical galaxy. 

At a redshift of $z = 0.297$, E1821 is one of the most optically luminous broad-line quasars in the local universe ($M_V = -27.1$, \citet{Hutchings1991Optical643}). It's host galaxy is a giant elliptical ($R_{1/2}=18.9$\,kpc,  $M_V=-24.3$; \citet{Floyd2004TheQuasars, Kim2017StellarNuclei}) which is located in a massive ($> 10^{15}$\,M$_\odot$), dynamically relaxed cluster \citep{Lacy1992TheE1821+643,Boschin2018Multi-objectHalo}. X-ray observations show that the cluster is a strong CCC with a short central cooling time of $\sim 1$\,Gyr and a temperature change from the outer to central regions of $9.0\pm1.5$ to $1.3\pm0.2$ keV \citep{Russell2010}. 

E1821 is therefore one of very few optically powerful quasars known to reside in the brightest cluster galaxy of a rich CCC, which more usually host moderately powerful radio galaxies.  Although E1821 is classified as a radio-quiet quasar (RQQ), it is associated with an extended ($\sim 250$\,kpc) FR\,I-like radio source oriented in the North-West (NW) and South-East (SE) directions \citep{Blundell2001TheStructure}, which appears to be produced by arcsecond-scale twin jets \citep{Blundell1996ARQQ}. The jet axis is oriented NE-SW within $\sim 0.5\arcsec$ of the core, but beyond this distance the SW jet bends by $\sim80^\circ$ around to the SE, becoming approximately parallel to the larger scale radio source, perhaps indicating precession of the radio jet axis.

Estimates of the central black hole mass range from $1-6\times 10^9\,M_\odot$ \citep{Dasyra2011AARCHIVE, Reynolds2014THEMEDIUM, Shapovalova2016FIRSTCONTINUUM, Kim2017StellarNuclei}.  It seems possible that the quasar's enormous bolometric luminosity ($L_{bol}\sim 2\times 10^{47}$\,erg\,s$^{-1}$) could be fuelled by Bondi accretion from the cooling flow itself \citep{Reynolds2014THEMEDIUM}.  However, there is substantial evidence that E1821 is in the late stages of a tidal interaction or merger  \citep{Fried1998,Blundell2001TheStructure,Aravena2011,Hutchings1991Optical643}. Moreover, E1821 is known to be a hyper-luminous infrared galaxy, with $\log L_{IR} = 13.1\,L_{\odot}$.  \citet{Farrah2002Sub-millimetreGalaxies} and \citet{Floyd2004TheQuasars} note that the quasar nucleus is unusually red, which implies a large amount of dust. Modelling of the spectral energy distribution indicates that star formation accounts for around 50\% of the total IR luminosity ($\sim 5\times 10^{12}$\,L$_{\odot}$), corresponding to a very high star formation rate (SFR) $\sim 1000$\,M$_\odot$\,yr$^{-1}$ \citep{Farrah2002Sub-millimetreGalaxies,Aravena2011}.  These properties suggest that E1821 is being observed late in the post-merger blowout phase, when an optically luminous quasar has emerged, but is still surrounded by a powerful starburst and a large amount of dust \citep{Aravena2011}.

\begin{figure}
    \centering
    \includegraphics[width=\columnwidth]{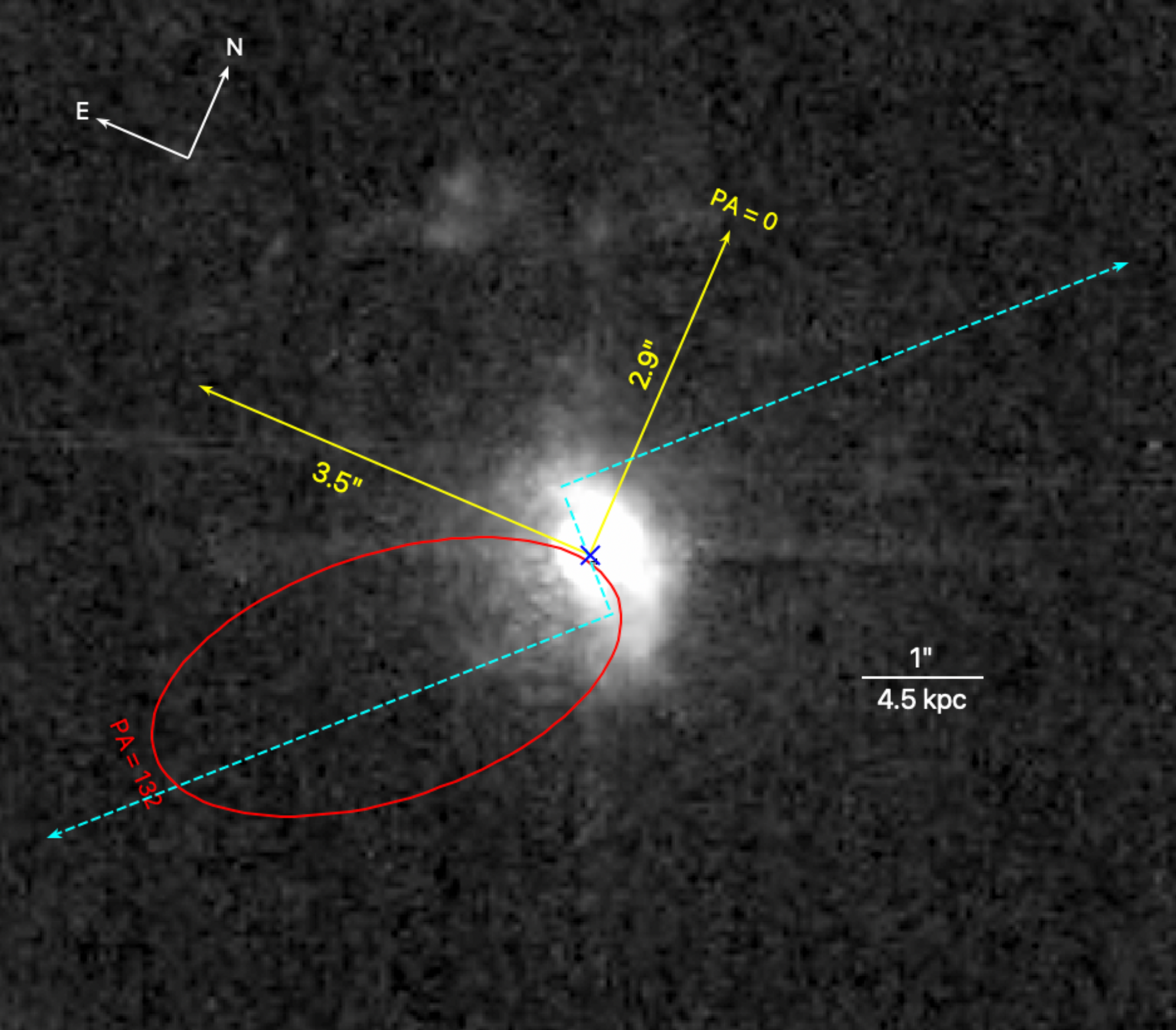}
    \caption{HST ACS narrow-band image of the extended [OIII]$\lambda\lambda$4959,5007 emission around the nucleus of E1821+643. The image was obtained with the FR656N ramp filter and the quasar point source has been subtracted using an image of a white dwarf as a PSF template \citep{Jadhav2021TheRecoil}. The dashed cyan lines indicate the approximate directions of the radio jets from \citet{Aravena2011} and \citet{Blundell2001TheStructure}. The yellow arrows show the distances to a faint arc of line emission that stretches from the North (PA $= 0^\circ$) to East (PA $=90^\circ$).  The red ellipse marks the approximate location, centered 2\arcsec SE, of the CO detection reported by \citet{Aravena2011}.  The blue X marks the location of the quasar nucleus (i.e., the peak of the point-spread function).}
    \label{hst}
\end{figure}

There is also evidence of ongoing tidal interactions.  \citet{Aravena2011} reported a CO detection, centered 2\arcsec\ SE of the quasar nucleus.  The detection, indicated in Figure \ref{hst}, has a roughly elliptical morphology and has a major axis of 4\arcsec.  They suggest this molecular gas has a large enough dynamical mass to be a separate gas-rich galaxy, which is interacting with E1821.  

It has also been proposed that the SMBH in E1821 is undergoing gravitational recoil, which, in itself is evidence for a previous major merger.  During galaxy mergers, dynamical friction will drive the SMBHs to spiral inward and form a SMBH binary \citep{Begelman1980MassiveNuclei}.  The release of gravitational waves bring the SMBHs closer until they finally coalesce.  Conservation of linear momentum results in a recoil kick to the merged SMBH remnant, causing the black hole and it’s bound material to move away from the center of the host galaxy (e.g., \citet{Bekenstein1973Gravitational-RadiationHoles, Favata2004HOWREVISITED}).  Numerical relativity simulations show that the kick velocity can reach $\sim 5000$\,km\,s$^{-1}$ (\citet{Campanelli2006}; \citet{Campanelli2007MaximumRecoil}; \citet{Lousto2012GravitationalBinaries}).  If it has an accretion disk, a kicked SMBH might appear as an offset AGN \citep{Blecha2016RecoilingAlignment}. For kick velocities $\gtrsim 1000$\,km\,s$^{-1}$, the broad line region (BLR) would be largely retained and move with the SMBH, while the narrow line region (NLR) is left behind, producing a velocity shift between the broad and narrow emission lines in the optical spectrum (\citet{Loeb2007ObservableMerger}; \citet{Bonning2007RECOILINGQUASARS}).  \citet{Robinson2010} performed a spectropolarmetric analysis of the broad H$\alpha$ line in E1821 and found large ($\sim 1000$\,km\,s$^{-1}$) velocity shifts in both total and polarized light.  Using a simple scattering model, they inferred that the gas bound to the SMBH i.e., that producing the broad lines, has a velocity shift $\sim 2100$\,km\,s$^{-1}$ relative to the  to the systemic velocity of the host galaxy as indicated by the narrow lines.  This suggests that the SMBH in E1821 is undergoing gravitational recoil following coalescence of a progenitor binary. 

If this is the case, there should also be a spatial offset between the gas emitting the broad and narrow lines, i.e., between the nucleus and the NLR.
\citet{Jadhav2021TheRecoil} looked for this using a combination of HST optical images and spectroastrometry. The HST image shown in Figure \ref{hst} reveals spatially resolved [OIII]$\lambda\lambda 4959,5007$ emission on kpc scales. The extended emission has an asymmetric distribution, bulging out to the North-West (NW) and spanning nearly 180$^\circ$. The [OIII] lines exhibit a spectroastrometric offset in the same direction, relative to the broad Balmer lines and quasar continuum.  This implies that the quasar nucleus is displaced by $\sim 580$\,pc to the South-East (SE), with respect to the [OIII] emission, consistent with the expected gravitational recoil offset.

Although E1821 is known to have a large extended emission line region (EELR), covering a few $\times 10$\,kpc \citep{Fried1998}, it has not been well studied. In this paper, we use the long-slit spectra obtained by \citet{Jadhav2021TheRecoil} to ``look" for the kinematic signatures of a quasar wind and more generally, to study the kinematics and ionization of the extended emission line gas around E1821.  The ionized gas could be gas expelled in the quasar's blowout phase, from a tidal tail, or filaments condensing from the cluster cooling flow.  In Section~\ref{sec:obs} we describe the observations, the extracted 1-D spectra, and the line fitting procedure.  In  Section~\ref{sec:results} we present the results derived from our line fits, including the variations of the kinematics and emission line ratios along the North-South (N-S) and East-West (E-W) axes, centered on the nucleus. In Section \ref{sec:discussion}, we discuss the kinematics and ionization of the gas and its possible origin.  In Section~\ref{sec:conc}, we present our conclusions.  We assume standard cosmological parameters, $H_o$ = 67.8 km\,s$^{-1}$\,Mpc$^{-1}$, $\omega_m$ = 0.308, and $\omega_\Lambda$ = 0.692. The angular scale is 4.556\,kpc\,arcesc$^{-1}$.  

\section{Observations}
\label{sec:obs} 

\begin{figure}
    \centering
    \includegraphics[width=\columnwidth]{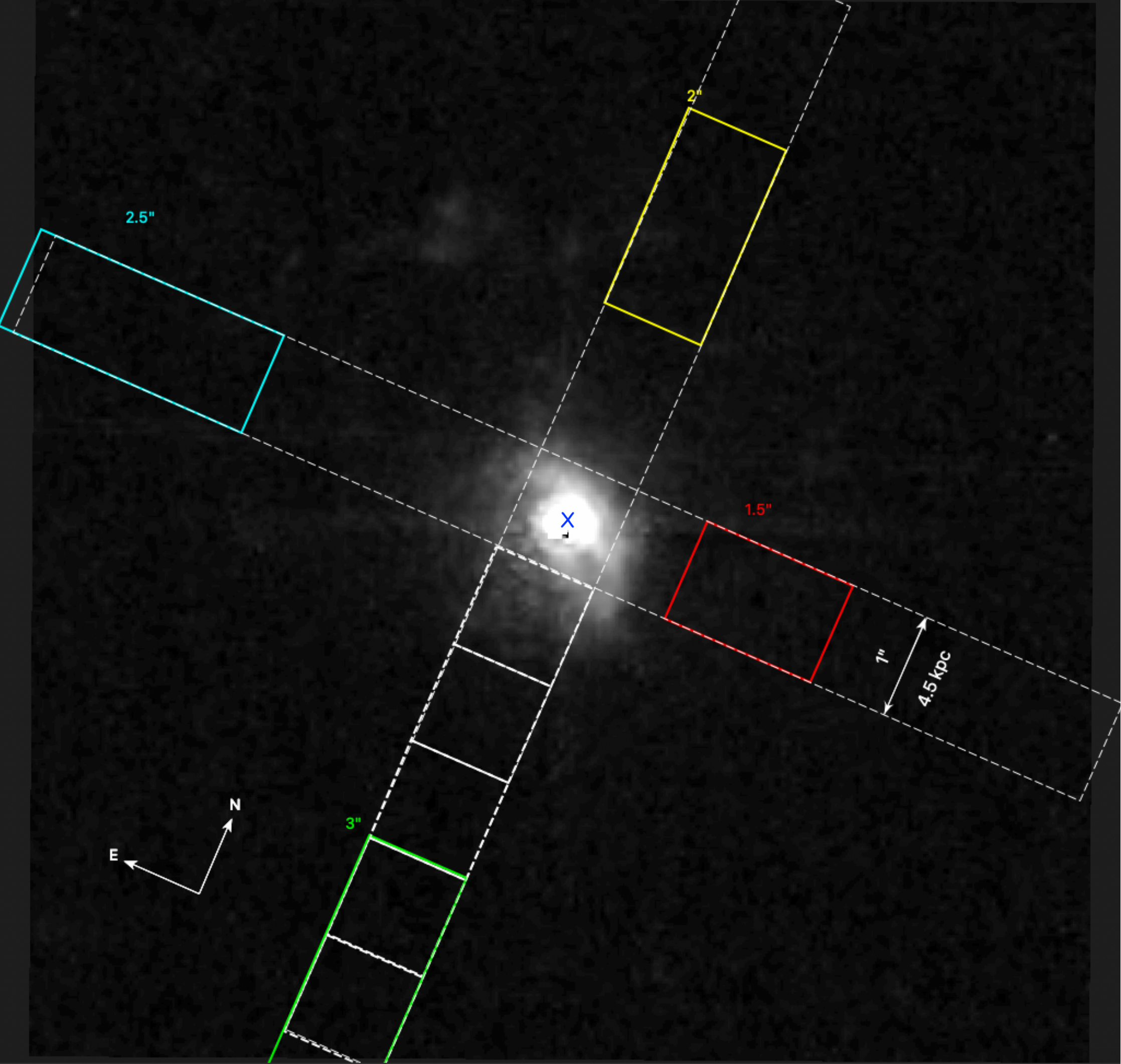}
    \caption{Spectrograph long-slit apertures (dashed, white rectangles) superimposed on the HST ACS image as shown in Figure \ref{hst}.  The slits are 1\arcsec~(4.5\,kpc) in width and 330\arcsec\ long, but only the central 10\arcsec~(45\,kpc) is shown in each direction. The blue X marks the nucleus center.  1-D spectra were extracted from apertures of 1\arcsec (white boxes), 1.5\arcsec (red), 2\arcsec (yellow), 2.5\arcsec (blue), and 3\arcsec (green) along the slits in each direction. For clarity, these apertures are shown only in one direction. The locations and sizes of the extraction apertures are listed in Table \ref{tab:size}.}  
    \label{fig:hst_obs}
\end{figure}

% Brief description: telescope, observing set up 
Long-slit spectra of E1821 were obtained with the Gemini Multi-Object Spectrograph (GMOS) of the Gemini North telescope in Hawaii on the nights of 20100424 (science) and 20100323 (flux calibrations). At the time of the observations, the GMOS camera utilized an array of 3 EEV CCDs, giving a pixel scale of 0.0727\arcsec/pixel. During the observations, the seeing conditions varied between full-width, half-max (FWHM) $= 0.8 - 1.1\arcsec$.  The spectra were obtained using the R400 grating and a 1\arcsec-wide slit centered on the quasar nucleus. This configuration produced a spectral resolution of $7.98\pm0.07$\AA.  The main purpose of the observations was to perform spectroastrometric measurements of the broad and narrow lines and to facilitate this, two sets of spectra were obtained, with the slit orientated in position angles PA = 0$^\circ$ (N), 90$^\circ$ (E), 180$^\circ$ (S) and 270$^\circ$ (W).  The spectroastrometry analysis is described by \citet{Jadhav2021TheRecoil}.  These authors also present HST Advanced Camera for Surveys (ACS) images to map the extended [OIII]$\lambda\lambda$4959,5007 emission around the nucleus. The spectroscopic slit positions are shown over-plotted on a point source subtracted [OIII] image in Figure \ref{fig:hst_obs}. 

\subsection{Data reduction}
\label{sec:red} 
% Brief description of standard long-slit reductions; extraction of 1-D spectra; removal of quasar light from PSF 
The data reduction procedures are described in \citet{Jadhav2021TheRecoil}.  Briefly, the \textsc{IRAF Gemini: GMOS} package was used to debias, flat field, sky subtract, and stack the 2D spectra, as well as performing wavelength and flux calibrations. The 2D spectra obtained at PA = 0$^\circ$ and 180$^\circ$ were combined to form a single 2D spectrum along the North-South (N-S) direction with a total exposure time of 1800\,s. Similarly, the PA = 90$^\circ$ and 270$^\circ$ spectra were combined to form a single spectrum in the East-West (E-W) direction with a total exposure time of 2400\,s.  In addition, all the 2D spectra obtained at both PA's were combined to form a single spectrum with a total exposure time of 4200\,s.

\begin{figure}
    \centering
    \includegraphics[width=\columnwidth]{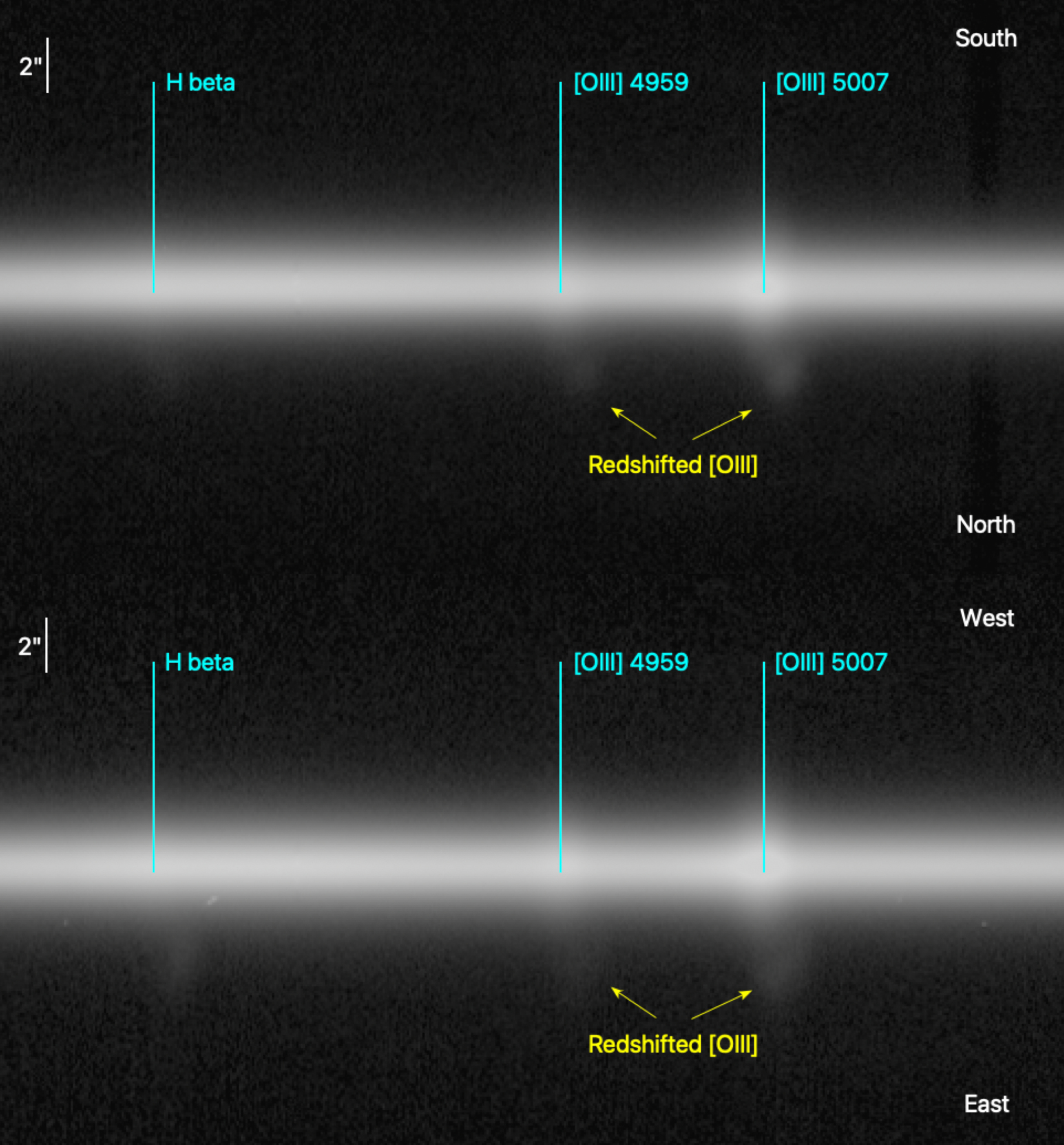}
    \caption{Segments of the combined 2D spectrum along the N-S (top) and E-W (bottom) directions. The extended emission in the $H\beta\lambda$4861 and [OIII]$\lambda\lambda$4959,5007 emission lines can be clearly seen. Along the spatial axis, 1\arcsec~= 12.5 pixels.} 
    \label{2dspec}
\end{figure}

\begin{figure*}
    \centering
    \includegraphics[width=0.9\textwidth]{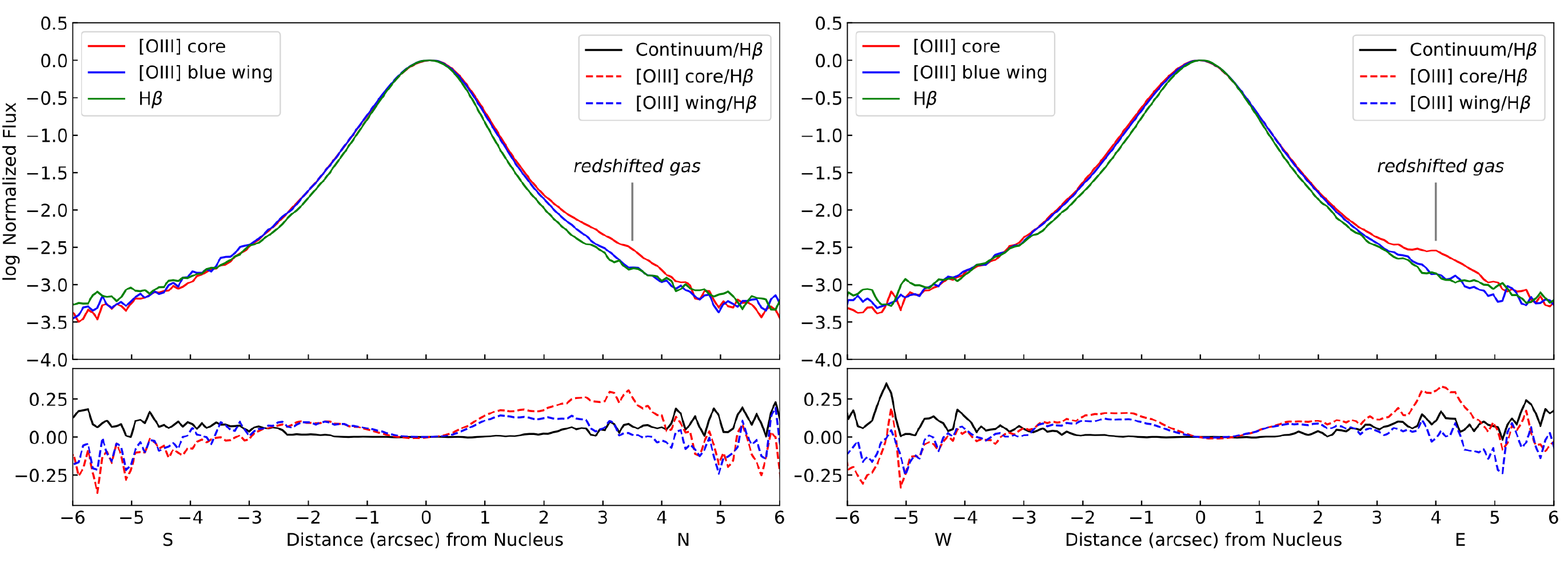}
    \caption{Spatial profiles extracted from the NS (left) and EW (right) 2D spectra shown in Figure \ref{2dspec}. The profiles were extracted from slices with a width of 5 pixels ($\sim3$\AA) and sample the core (red line) and blue wing (blue line) of [OIII]$\lambda 5007$\AA, and the central peak of the broad H$\beta$ line profile (green line).  The slices are centered on the wavelengths of the gaussian components that were used to model these components in the 1D spectrum extracted from the nucleus, at 6493.7\AA, 6487.7\AA, and 6334.6\AA, respectively (see Figure \ref{fig:Nuc_Hb}). Each profile is normalized to its peak flux. The lower panels show the ratios, also in the log, [OIII] core/H$\beta$ (dashed red line), [OIII] blue wing/H$\beta$ (dashed blue line) and for comparison, the continuum/H$\beta$ ratio (black line). The continuum spatial profile is the average of profiles extracted from line free regions either side of the broad H$\beta$ line at wavelengths of 6190.6\AA\ (slice width 40 pixels) and 6652.1\AA\ (12 pixels).} 
    \label{fig:spatial}
\end{figure*}

Segments of the combined N-S and E-W 2D spectra, including the H$\beta\lambda$4861 and [OIII]$\lambda\lambda$4959,5007 lines, are shown in Figure~\ref{2dspec}. The narrow line emission is clearly extended on all sides of the nucleus, but is brightest and extends the furthest to the N and E. The emission in these directions separates into two distinct components, one extending $2-3$\arcsec from the nucleus and a second, which is further out ($\sim 3-4$\arcsec N and $4-5$\arcsec E, respectively) and also redshifted.  In Figure \ref{fig:spatial} we show the spatial profiles centered on the core and blue wing components of [OIII]$\lambda 5007$\AA, and the central peak of the broad H$\beta$ line (see Section~\ref{sec:nucleus} and Figure \ref{fig:Nuc_Hb}), normalized to their respective peak fluxes. The profiles centered on the [OIII] core and wing components are slightly broader than that of the broad H$\beta$ line beyond distances $\sim 1\arcsec$. This is clearly seen in the lower panels of Figure~\ref{fig:spatial}, which show the ratios of the spatial profile ratios of, respectively, the [OIII] core and blue wing components to the broad H$\beta$, and for comparison, the ratio of the average spatial profile of two continuum regions to broad H$\beta$. Since the continuum is dominated by the quasar, we expect the continuum and broad H$\beta$ to have the same spatial profiles and, indeed, the continuum/H$\beta$ ratio is $\approx1.0$ within 4\arcsec~of the nucleus.  In contrast, the core and wing [OIII]/H$\beta$ ratios both deviate from 1.0 beyond the central 1\arcsec~($\sim$the seeing disk), reaching values up to $1.5-1.6$ to the North and West, indicating that [OIII] emission is resolved. The bulges that appear at approximately 2\arcsec~ - 5\arcsec~E and 2\arcsec~- 4\arcsec~N in the [OIII] core profile correspond to the redshifted emission visible in the 2D spectra (Figure~\ref{2dspec}).  The spatial profiles show that the blue wing component is approximately co-spatial with the core component within the inner 2\arcsec and is present in all directions.    

From these 2D spectra, 1D spectra were extracted from 1\arcsec\ apertures, beginning with the nucleus itself, extracted from the combined 2D spectrum described above, then at 1\arcsec\ distance increments in each direction.  The 1\arcsec\ apertures are well matched to the seeing. However, since the flux decreases rapidly with distance from the nucleus, larger extraction apertures were used in addition to the standard 1\arcsec aperture.  The extraction apertures are shown overlaid on the [OIII] image in Figure~\ref{fig:hst_obs} and Table \ref{tab:size} lists the directions, distances, and aperture sizes used for the 1D spectra. The spectra extracted from the 1\arcsec apertures centered on the nucleus exhibit the strong, broad Balmer lines H$\gamma\lambda$4340, H$\beta\lambda$4861, and H$\alpha\lambda$6562, and the blue continuum that is characteristic of quasars. The quasar itself is unresolved, however, from  Figures~\ref{2dspec} and~\ref{fig:spatial}, we can see that light from the nucleus is spread over the central $\sim 4$\arcsec. Therefore, in the  extraction apertures offset from nucleus there is substantial contamination by light in the wings of the quasar point spread function (PSF). To remove this contamination, we fitted and subtracted the narrow lines in the spectrum of the nucleus and then subtracted a scaled version of this spectrum (containing the quasar continuum and broad lines) from the offset spectra (see Section~\ref{sec:nucleus}, for details of the fitting method).  

The HST/ACS images were obtained in the narrow band FR656N ramp filter, set to 6460\AA, to map the redshifted [OIII]$\lambda\lambda$4959,5007 emission (Proposal ID 13385, PI A.Robinson). As described in \citet{Jadhav2021TheRecoil}, the bright quasar PSF was modelled using both \textsc{GALFIT} \citep{Peng2010DetailedModels} and a white-dwarf standard star and then subtracted. The [OIII] image, after subtraction of the the quasar PSF, is shown in Figures \ref{hst} and \ref{fig:hst_obs}. The extended [OIII] emission has an elongated, asymmetric morphology, which is approximately 1.6\arcsec ~long, 1\arcsec ~wide, and off-center with respect to the nucleus. Inspection of the [OIII] image shows that the long axis of this structure is oriented approximately NE-SW, and bulges outwards NW of the nucleus.  There is also a faint arc-like structure about 3\arcsec ~NE from the nucleus.  Unfortunately, most of this feature falls between our slit positions (Figure \ref{fig:hst_obs}).

\begin{table}
 \caption{Extraction regions for the 1D spectra. The columns list the label we will use to identify the direction, distance from the nucleus, and size of the extraction aperture, respectively. }
\begin{tabular}{ |l|c|c| } 
 Spectrum ID & Distance & Aperture\\ 
 \hline
 Nuc-0 & 0" & 1"\\
 N/S/E/W-1 & 1" & 1"\\ 
 N/S/E/W-2 & 2" & 1"\\ 
 N/S/E/W-2-lg & 2" & 1.5"\\ 
 N/S/E/W-3 & 3" & 1"\\ 
 N/S/E/W-3-lg & 3" & 2"\\ 
 N/S/E/W-4 & 4" & 1"\\ 
 N/S/E/W-4-lg & 4" & 2.5"\\ 
 E/W-5 & 5" & 1" \\
N/S-5 & 5" & 2"\\ 
 \hline
\end{tabular}
\label{tab:size}
\end{table}

\subsection{Emission line fitting}
\label{sec:fitting}

We fit the emission lines in the 1D spectra with gaussian components using the \textsc{python} fitting package \textsc{pyspeckit}, to measure the flux, velocity, and velocity dispersion at each position.  To reduce the number of free parameters, a number of constraints based on nebular physics were applied to the fits to the forbidden lines.  For both components, the amplitude of the [OIII]$\lambda 4959$ line is scaled to that of the  stronger [OIII]$\lambda 5007$ line in a 1:3 ratio \citep{Osterbrock2005}. The wavelength and width are similarly scaled to [OIII]$\lambda 5007$ by the ratio of wavelengths, $4958.9/5006.9$. The weaker lines of the [OI]$\lambda\lambda$6300,6363 and [NII]$\lambda\lambda$6548,6583 doublets were similarly tied in amplitude, width, and wavelength to their respective strong components.  The two [SII]$\lambda\lambda$6717,6731 lines were also tied in wavelength and width. 

\subsubsection{Nucleus}\label{sec:nucleus}

The focus of this paper is the EELR surrounding the nucleus. However, the narrow lines were also fitted in the spectrum of the nucleus, Nuc-0, for comparison with the EELR, and to create a template for removal of the quasar PSF from the offset spectra. In order to achieve satisfactory fits to the narrow lines, it was necessary to flatten the spectrum by fitting and removing the underlying continuum. Two components were used, a power-law $f_\lambda \propto \lambda^{-2.33}$, representing the accretion disk spectrum and a low-order polynomial, to account for artifacts in the flux calibration sensitivity curve. This combination was fitted to several line free regions in the continuum (5775--5782, 6177--6204, 6648--6656, 7342--7460, 8011--8102 and 8958--8994\AA). After subtraction of the fitted continuum, the broad and narrow lines were separately fitted in the wavelength ranges 6208--6801\AA\ and 8074--8991\AA, which encompass the H$\beta$-[OIII] and the H$\alpha$-[NII] regions, respectively.

\begin{figure}
    \centering
    \includegraphics[width=\columnwidth]{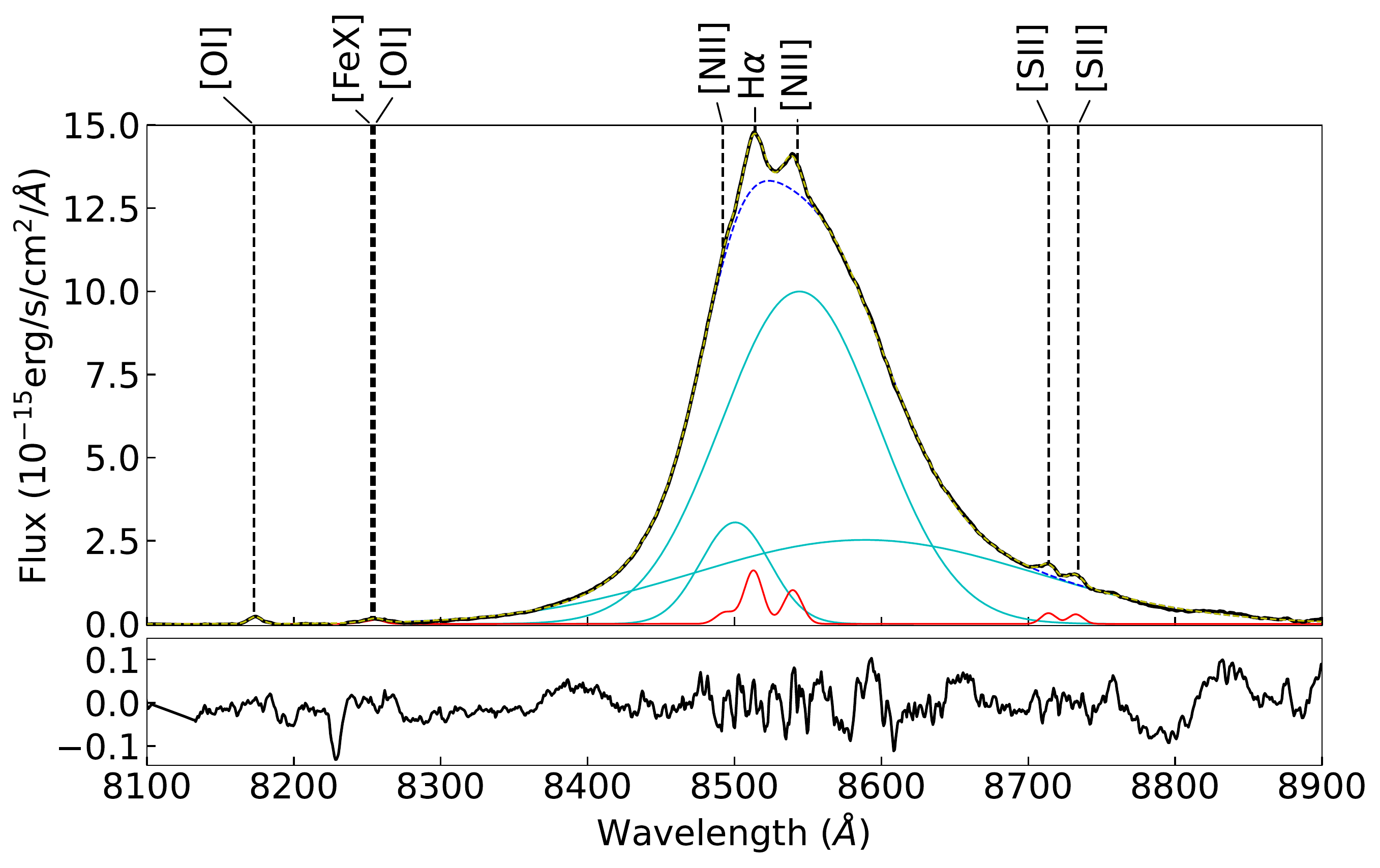}
    \caption{Fit to H$\alpha$ and adjacent lines in the nucleus. The broad H$\alpha$ line profile is modelled with 3 components, shown in cyan and the overall fit is the dashed blue curve.} The fit to the narrow lines (narrow H$\alpha$, [OI]$\lambda\lambda$6300,6363, [NII]$\lambda\lambda$6548,6583 and [SII]$\lambda\lambda$6717,6731 is shown in red. The data is in black, and the overall fit is the dashed yellow line.  The residuals of the overall fit are plotted in the lower panel.
    \label{fig:Nuc_Ha}
\end{figure}

In the H$\alpha$ segment, [OI]$\lambda\lambda$6300,6363, [NII]$\lambda\lambda$6548,6583 and [SII]$\lambda\lambda$6717,6731 were fitted with single gaussian components, applying the constraints described above. The [FeX]$\lambda$6374 and the narrow H$\alpha$ lines were also fitted by a single gaussian, but no constraints were applied. In order to achieve a good fit, it was necessary to model the broad H$\alpha$ line with 3 components, a strong ``core'' component, a broader, redshifted ``base'', and a weaker blue-shifted ``shoulder'' component, as shown in Figure~\ref{fig:Nuc_Ha}. 

\begin{figure}
    \centering
    \includegraphics[width=\columnwidth]{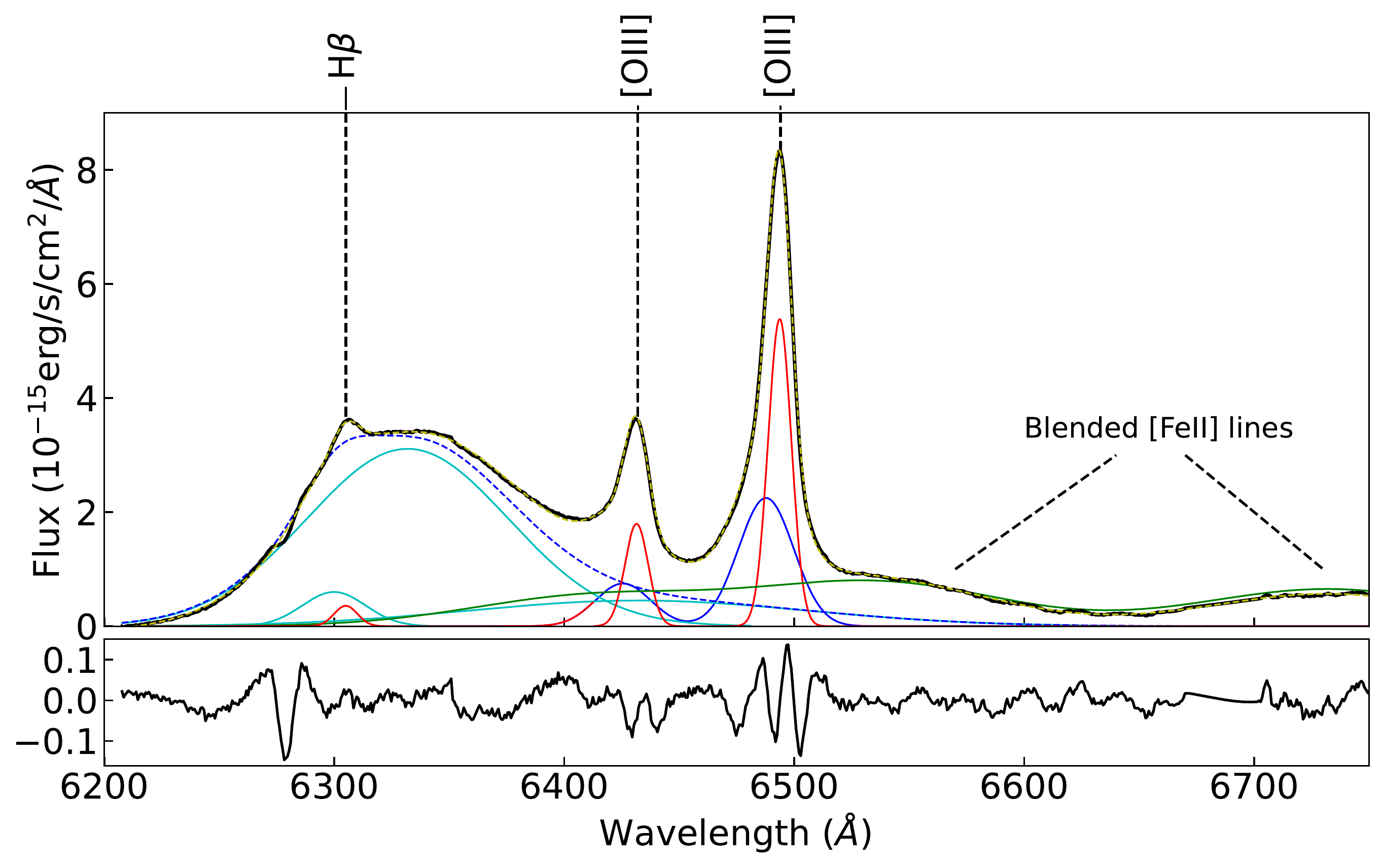}
    \caption{Fit to the H$\beta$ and [OIII]$\lambda\lambda$4959,5007 lines in the nucleus. The broad H$\beta$ line is modelled with 3 broad components (cyan) and the overall fit is the dashed blue curve.  The narrow H$\beta$ is modelled with a single component (red). The} [OIII] lines have two components, representing the core (red) and a blue wing (blue). The broad components underlying the [OIII] lines and around 6700-6800\AA~ are blended FeII lines (green). As in Figure \ref{fig:Nuc_Ha}, the black line is the data and the overall fit is the yellow dashed line.
    \label{fig:Nuc_Hb}
\end{figure}

In the H$\beta$ region, the [OIII]$\lambda\lambda$4959,5007 lines are much stronger than the other narrow lines and are clearly asymmetric, exhibiting blue wings. The [OIII] lines were therefore modelled with two gaussian components representing the ``core'' and the the ``blue wing'', respectively. For both components, [OIII]$\lambda 4959$ was tied to [OIII]$\lambda 5007$ as described above. The narrow H$\beta$ line was fitted with a single gaussian component, with no constraints applied. The broad H$\beta$ line profile was modelled with three components as for H$\alpha$. Several broad FeII lines are also present in this region, some of which underlie the [OIII] lines, most notably the M42 multiplet lines at rest wavelengths 4923.9, 5018.4 and 5169.0\AA. These lines were each modelled with a single, broad gaussian component, with the intensity, width and wavelength of the components representing the 4923.9 and 5169.0\AA\ lines being tied to the 5018.4\AA\ line. The relative intensities of the 4923.9 and 5169.0\AA\ lines with respect to the 5018.4\AA\ line were taken from \citet{Kovacevic2010ANALYSISSPECTRA} (0.693 and 0.854, respectively, assuming a gas temperature T = 10,000\,K). An additional broad gaussian component was included in the fit to model the rest of the heavily blended FeII lines that form the ``red bump'' at rest wavelengths $\sim 5150-5400$\AA. The resulting fit is shown in Figure \ref{fig:Nuc_Hb}. 

After performing the fits, the gaussian components representing the narrow lines were subtracted from the spectrum to create a template containing only the unresolved quasar continuum and broad emission lines (the contribution from the host galaxy's stellar continuum is negligible -- \citet{Jadhav2021TheRecoil}). Scaled versions of this template spectrum were then used to remove the quasar PSF light from the offset spectra.

It is likely that the blue wing seen in [OIII]$\lambda\lambda$4959,5007 is also present in the other narrow lines. However, these lines are much weaker than both of the [OIII] lines and the narrow H$\alpha$, [NII] and H$\beta$ lines are also heavily blended with the corresponding broad lines. Blue wings are not apparent from close visual inspection of the spectra. Nevertheless, we also performed fits to the H$\alpha$ and H$\beta$ regions in which blue wing components were included for the narrow H$\alpha$ and [NII] lines (Figure \ref{fig:Nuc_Hawing}) and the narrow H$\beta$ line (Figure \ref{fig:Nuc_Hbwing}). The blue wing components were tied in width and velocity shift to the corresponding [OIII] blue wing component, but the amplitudes were left as free parameters. In the H$\alpha$ region, blue wing components were not included for the [OI] and [SII] lines, since these components were completely suppressed in trial fits. Including the blue wing components to the narrow H$\alpha$, H$\beta$ and [NII] lines did not significantly improve the fits. For the H$\alpha$ region, $\chi^2 = 2.19~(2.21)$ and for the H$\beta$ region $\chi^2 = 0.80~(0.91)$, with (without) the wing components. Furthermore, the amplitudes of the narrow H$\alpha$ and [NII] blue wing components are poorly constrained, leading to large uncertainties in the fluxes. Overall, therefore, the strengths of the blue wings in lines other than [OIII]$\lambda\lambda$4959,5007 are not reliably constrained by the fits.

In the following analysis, the kinematics and emission line fluxes of the nucleus region, Nuc-0, were derived from the fits described above, in which the blue wing component is included only for the [OIII] lines. In the BPT diagrams described in Section~\ref{sec:ionization}, the [OIII]/H$\beta$ for the Nuc-0 point is calculated using the flux of the core component of [OIII]$\lambda$5007.

\subsubsection{Extended emission line region}\label{sec:extended}

Although most of the contaminating PSF light was removed from the offset spectra by scaling and subtracting the template spectrum of the quasar nucleus, some residual light remained, manifesting as a slight increase to the blue end of the spectra. Prior to line fitting, therefore, we fitted and subtracted a power-law baseline to flatten the spectra. 

With the exception of [OIII]$\lambda\lambda$4959,5007 (see below), all lines were fitted with a single gaussian component, applying the constraints described at the beginning of this Section.
However, the [SII] lines are relatively weak in many of the spectra and also moderately blended. For these reasons, and as it is reasonable to assume that the [SII] and [NII] doublets are largely emitted by the same gas, the [SII]$\lambda$6717 line was also tied in width to the [NII]$\lambda$6583 line. 

A simple Monte-Carlo (MC) procedure was employed to estimate the uncertainties in the quantities derived from the fits. 10 realizations of each spectrum were generated by adding a random offset value to the measured flux in each wavelength bin. The offsets were drawn from a normal distribution, with zero mean and standard deviation ($\sigma$) set to the uncertainty in the measured flux. For each spectrum, the realizations were fitted separately until 10 good fits had been obtained. The most probable value for each of the emission line parameters and its uncertainty were taken to be, respectively, the mean and the standard deviation derived from the fit iterations.

As explained in Section \ref{sec:red} and shown in Table \ref{tab:size}, we extracted 1D spectra at 1\arcsec\ intervals in distance using a standard 1\arcsec~ extraction aperture. However, since the flux decreases with distance (the SNR in the continuum decreases from $\sim5$ to $\sim2$), larger aperture sizes were also used to obtain a better signal for the weaker lines. For instance, at N-4 and N-5 the larger apertures were necessary to obtain accurate velocity measurements and to measure various emission line flux ratios. However, it should be noted that the adjacent larger apertures overlap and so measurements made from the corresponding spectra should not be considered independent.     

\begin{figure}
    \centering
    \includegraphics[width=\columnwidth]{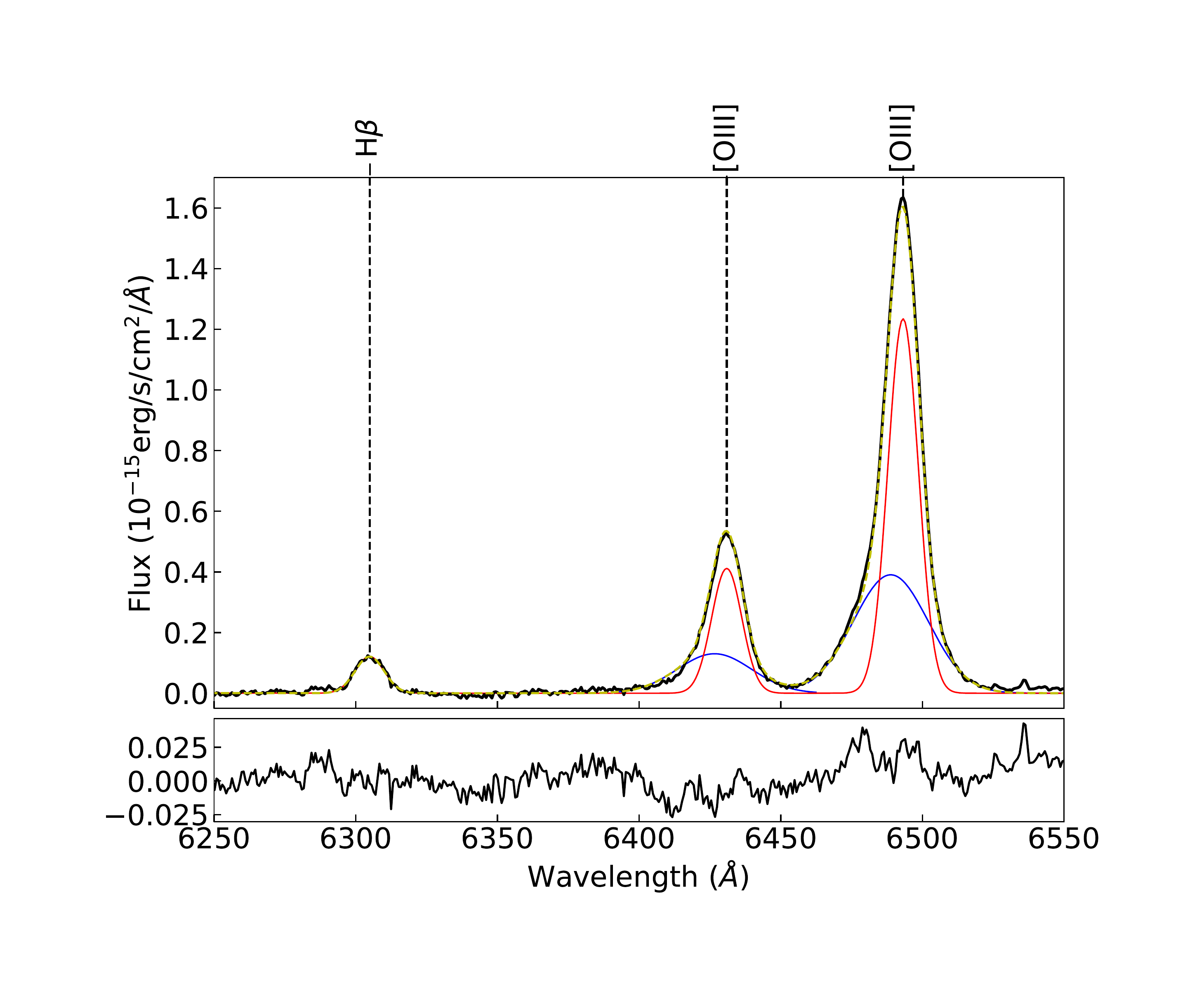}
    \caption{E-1 spectrum showing the fits to the H$\beta\lambda$4861 and [OIII]$\lambda\lambda$4959,5007 lines after subtraction of the scaled quasar PSF template spectrum (Section~\ref{sec:nucleus}). The data is the black line and the overall fit to the data is the yellow dashed line.  Each [OIII] line is modelled by a narrow `core' component (red ) and a broader, blueshifted `wing' component (blue).}
    \label{oiii}
\end{figure}

\begin{figure}
    \centering
    \includegraphics[width=\columnwidth]{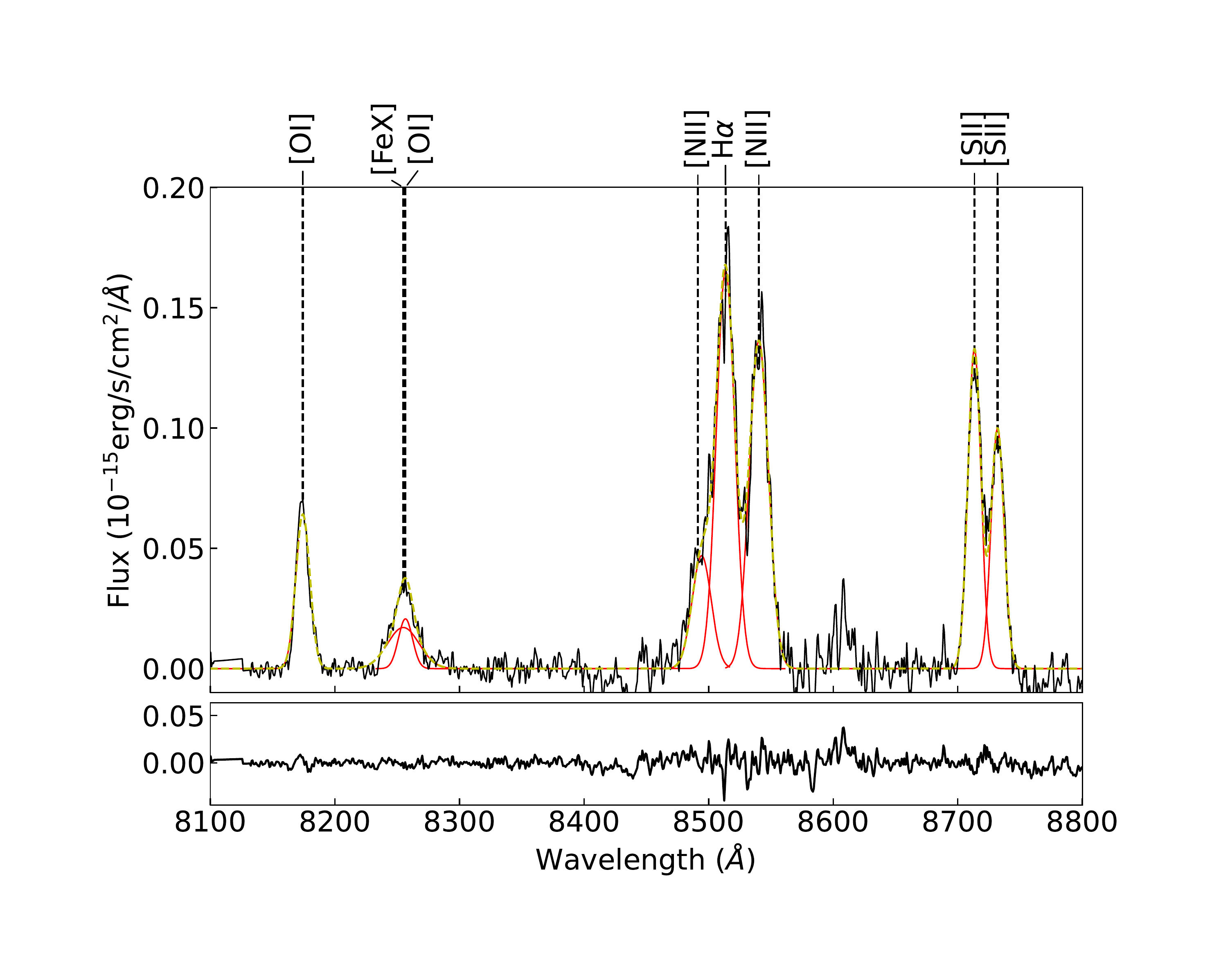}
    \caption{E-1 spectrum showing the fits to the [OI]$\lambda\lambda$6300,6364, H$\alpha\lambda$6563, [NII]$\lambda\lambda$6548,6583, and [SII]$\lambda\lambda$6716,6731 lines after subtraction of the scaled quasar PSF template spectrum (Section~\ref{sec:nucleus})}. The data is the black line and the gaussian fit components are shown in red.  The overall fit to the data is the yellow dashed line.
    \label{fig:halpha}
\end{figure}

An example of a fit to the H$\beta$ and [OIII]$\lambda\lambda$4959,5007 lines of spectrum E-1 is shown in Figure \ref{oiii}.  The narrow H$\beta$ line is fitted by a single gaussian component, without any constraints applied to its parameters.  The [OIII] lines are noticeably asymmetric in spectra extracted from the nucleus itself (Section~\ref{sec:nucleus})and from the regions within 3\arcsec of the nucleus, showing a pronounced blue wing. As the line profiles are not well represented by a single, narrow gaussian, we added a second, broad component, which is  blueshifted relative to the line peak. Hereafter, we will refer to the former feature as the [OIII] ``blue wing" component and to the latter as the [OIII] ``core" component.  Beginning at $\sim$3\arcsec from the nucleus the blue wing becomes indistinguishable from the noise and the inclusion of this component in the fit does not improve the $\chi^2$ value. Therefore, we omit the blue wing component and fit the [OIII] lines each with a single gaussian for extraction apertures at 3\arcsec\ and beyond, with the exception of the 2\arcsec\ apertures S-3-lg and W-3-lg. As in the nucleus (Section~\ref{sec:nucleus}), this broad, blueshifted wing is only clearly evident in [OIII]; similar components may be present in other lines but cannot be discerned in the spectra, so we model them with a single gaussian.  The the fit to H$\alpha$ and its neighboring lines at E-1 is shown in Figure \ref{fig:halpha}.  The complete spectrum from E-1 is shown in Figure \ref{fig:ewfull}. 

For each line, the velocity relative to the quasar nucleus was calculated,

\begin{equation}
\centering
v_{rel} = \frac{\lambda-\lambda_{nuc}}{\lambda_{nuc}}c
\end{equation}

where $\lambda$ and $\lambda_{nuc}$ are the wavelengths of a line in the EELR and quasar nucleus, respectively.  The velocity dispersion was obtained from the standard deviation of the fitted gaussians, which was converted to full-width half-max (FWHM) and corrected for the instrumental resolution, $7.98\pm0.07$\,\AA.  The line fluxes obtained from the fits were used to calculate the line ratios of [OIII]$\lambda$5007/H$\beta$, [OI]$\lambda$6300/H$\alpha$, [NII]$\lambda$6583/H$\alpha$, and [SII]$\lambda$6716,6731/H$\alpha$, which can give insight  into the nature of various possible ionization mechanisms, whether HII regions (star formation), shocks, or AGN photoionization.  

\section{Results}
\label{sec:results}

\subsection{Kinematics}\label{sec:kinematics}

\begin{figure*}
    \centering
    \includegraphics[width=\textwidth]{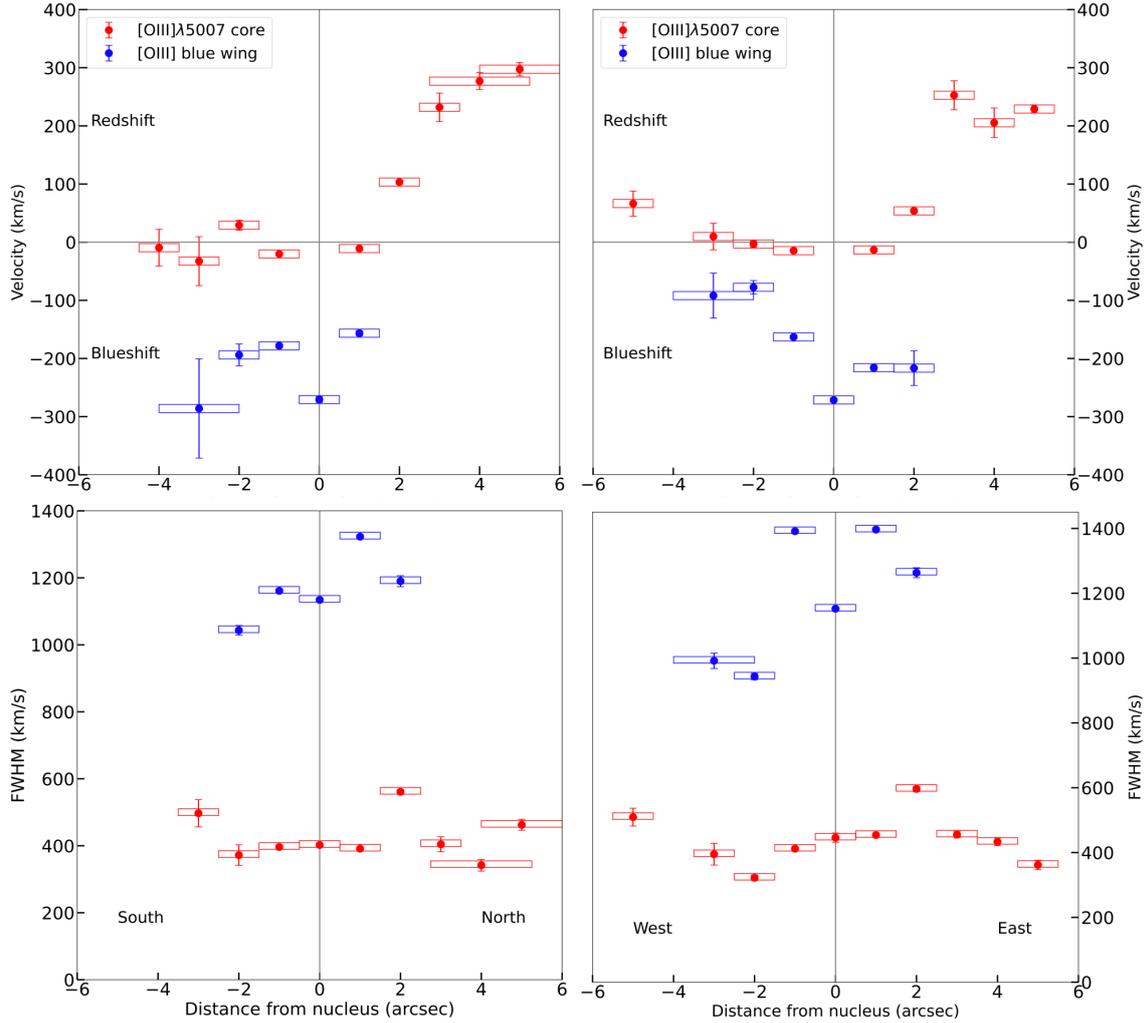}
    \caption{Relative velocities (\textit{top}) and velocity dispersion (\textit{bottom}) of the [OIII]$\lambda$5007 components plotted against distance from the nucleus in the N-S (\textit{left}) and E-W (\textit{right}) directions. The core and blue wing components are represented by red and blue points, respectively. The rectangles represent the extraction aperture size.  Negative distances are south of the nucleus and negative velocities correspond to blueshifts.  
    }
    \label{vel_oiii}
\end{figure*}

The velocities and velocity dispersions measured from the [OIII]$\lambda5007$ core component and its associated blue wing component are shown in Figure \ref{vel_oiii} in the N-S and E-W directions. Beginning at 2\arcsec~ to the North and East, the core component is redshifted relative to the nucleus. To the North, the velocity increases to 300\,km\,s$^{-1}$ at $\sim4-5$\arcsec\ (N-4-lg and N-5-lg). It should be noted that N-4-lg overlaps with N-3 and N-5-lg, nevertheless the trend is clear (larger extraction apertures were required to obtain sufficient SNR in this region). To the East, the velocity shift is slightly lower, $\sim 200-300$\,km\,s$^{-1}$ at 3-5\arcsec ~from the nucleus (E-3, E-4, and E-5).  To the South and West, the [OIII] core component velocity does not show any systematic variations with distance; the velocity relative to the nucleus fluctuates around zero by a few $\times 10$\,km\,s$^{-1}$.

The velocity dispersion of the [OIII] core component is fairly stable, with FWHM$\sim 400$\,km\,s$^{-1}$. However, in N-2 and E-2, where the velocity begins to increase, the FWHM is significantly higher, at $\sim 600$\,km\,s$^{-1}$. Therefore, the velocity dispersion increases across the boundary that separates the kinematically unperturbed gas from the redshifted gas.

The blue wing component is mainly confined close to the nucleus ($\leq2\arcsec$ N and S; $\leq2\sim3\arcsec$ E and W). At the nucleus (Nuc-0), this component is blueshifted by $\sim$300\,km\,s$^{-1}$ relative to the core [OIII] component. Within 1-2\arcsec~ N, S, and E the blueshift is roughly 200\,km\,s$^{-1}$ but decreases to $\lesssim 100$\,km\,s$^{-1}$ (or becomes undetectable) at larger distances. The FWHM of the blue wing component is significantly larger than that of the core component and has larger scatter, ranging between $\sim1000$ and $\sim1400$\,km\,s$^{-1}$. 

% Present velocity & velocity dispersion vs distance plots

\begin{figure*}
   \centering
    \includegraphics[width=\textwidth]{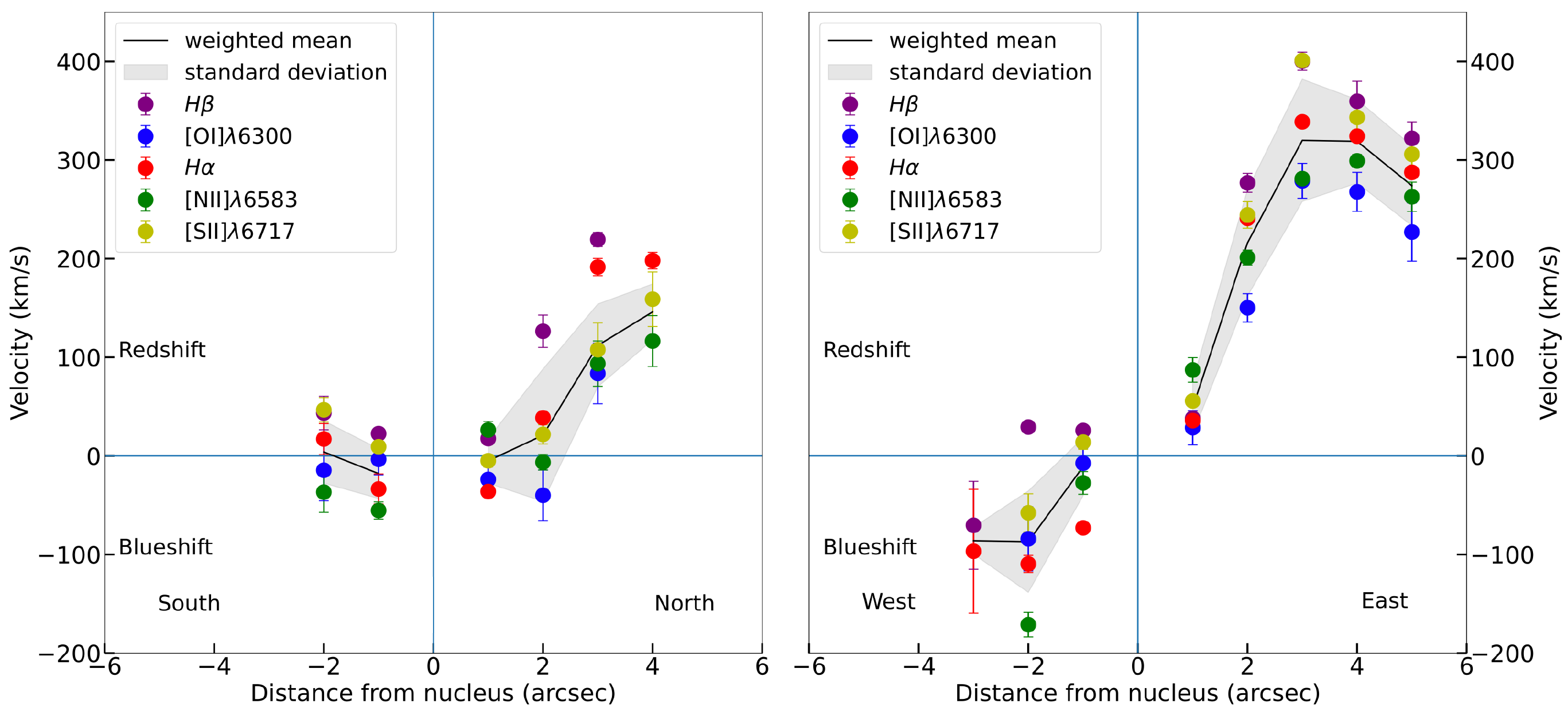}
    \caption{Relative velocities in the N-S (\textit{left}) and E-W (\textit{right}) directions for the narrow lines H$\beta$ (purple), [OI] (blue), H$\alpha$ (red), [NII] (green), and [SII] (yellow) versus distance from the nucleus. The black line represents the weighted mean for the listed lines at each 1\arcsec ~interval.  The shaded region is the weighted standard deviation.  The axes follow the same convention as in Figure \ref{vel_oiii}.  
    }
    \label{other_vel}
\end{figure*}

The velocities in the N-S and E-W directions for the H$\beta$, [OI], H$\alpha$, [NII], and [SII] lines are plotted in Figure \ref{other_vel}, which includes their weighted mean and weighted deviation.  These lines show behaviors in velocity shift very similar to those seen in the core [OIII] component.  To the South and the West there is no velocity shift, whereas, from 1-3\arcsec N, there is a sharp increase, which peaks at $\sim$300\,km\,s$^{-1}$.  There is a similar increase in velocity to the East, peaking at a redshift $\sim$300\,km\,s$^{-1}$ at 3-5\arcsec ~E.  The values and spatial variations of the FWHM are consistent with those of the core [OIII] component shown in Figure \ref{vel_oiii}.  

In summary, we can identify 3 distinct components in the kinematics of the EELR. There is the unperturbed, kinematically quiescent gas to the South and West with no velocity shift relative to the nucleus and a constant velocity dispersion (FWHM$\sim 400$\,km\,s$^{-1}$).  Secondly, there is a broad (FWHM$\sim 1000$\,km\,s$^{-1}$), blueshifted ($v_{rel}\sim -200$\,km\,s$^{-1}$) [OIII] wing component that is mainly detected in the nucleus and inner regions of the EELR ($\lesssim 2\arcsec$).  Lastly, there is a component that is redshifted by $v_{rel}\sim 200-300$\,km\,s$^{-1}$ relative to the nucleus, which is located at distances $\gtrsim 3\arcsec$ to the North and East. There is evidence for an increase in velocity dispersion (to FWHM$\sim 600$\,km\,s$^{-1}$) across the boundary between this component and the quiescent gas. 

\subsection{Ionization Structure}\label{sec:ionization}
Here we discuss the ionization mechanisms and the ionization structure of the EELR around E1821.  The line ratios for the N-S and E-W directions are presented in BPT diagnostic diagrams \citep[Figures \ref{ns_shock}-\ref{ew_ion};][]{Kewley2001OpticalGalaxies, Veilleux1987SpectralGalaxies, Baldwin1981CLASSIFICATIONOBJECTS} where, the symbol color represents distance from the nucleus, the shape distinguishes direction, and the point size indicates the extraction aperture size. The ``maximum starburst  line'' defined by \citet{Kewley2001OpticalGalaxies} is shown in red and separates HII regions associated with star formation from AGN. The nucleus and all of the EELR sampled by our spectra lie well within the AGN region (above the red line). The [OIII]$\lambda$5007/H$\beta$ ratio, a measure of the mean ionization, ranges between $\sim3$ and $\sim10$. 

As discussed in Section~\ref{sec:kinematics}, we distinguished 3 kinematically distinct components in the velocity field, the redshifted gas from  3 to 5\arcsec~N and 3 to 5\arcsec~E, the kinematically quiescent gas to the South and West, and the blue wing component of [OIII] in the nucleus and surrounding regions ($\lesssim 2\arcsec)$. However, N-5-lg and E-5 are not included since only the strong [OIII] line was measurable. The blue wing component is only reliably detected in [OIII]. Nevertheless, as described in Section~\ref{sec:nucleus}, we performed fits to the nucleus spectrum including a blue wing component to the H$\alpha$, H$\beta$, and [NII] narrow lines. We have included the flux ratio values for the core and blue wing components derived from these fits in the [OIII]/H$\beta$ to [NII]/H$\alpha$ BPT diagrams.  We will refer to these points as Nuc-core and Nuc-wing, respectively.    

In the N-S direction, the points representing the offset regions generally have lower [OIII]/H$\beta$ and higher values of the other line ratios than the nucleus (Nuc-0; with the exceptions of N-1 and S-1 in the [NII]/H$\alpha$ diagram).

The  redshifted regions, N-3-lg and N-4-lg, which spatially overlap, are close together (with overlapping error bars) and have a higher ionization state
%does not appear to be systematically different in its ionization state, 
compared to the quiescent gas.  Furthermore, the North and South 2\arcsec and 3\arcsec points are pretty consistently clustered together. In the E-W direction, the points representing the redshifted component, E-3-lg and E-4-lg, which also spatially overlap, cluster closely together with the E-2-lg point and have a lower value of the [OIII]/H$\beta$ ratio ($\sim 3$) than the nucleus (Nuc-0; $\sim 12$) and other regions ($\sim 6$).

The locations in the [OIII]/H$\beta$--[NII]/H$\alpha$ diagram of the nucleus core components, Nuc-core and Nuc-0, indicate relatively high ionization conditions, with both having [OIII]/H$\beta~(\gtrsim 10)$ and relatively low [NII]/H$\alpha$. On the other hand, the blue wing has relatively low [OIII]/H$\beta~(\sim 5)$ compared to the core components, and somewhat higher NII]/H$\alpha$ (although with a large uncertainty), suggesting that it arises in lower ionization gas than the core.

\subsubsection{Electron Density}
\label{sec:density}
The electron density (N$_e$) can be derived from the [SII] doublet intensity ratio \citep{Proxauf2014AstronomyNote}, however the relatively weak [SII] lines are partially blended in our spectra, especially at distances larger than 2\arcsec.  Weighted mean estimates for the gas density in the nucleus and the EELR from 1-2\arcsec in each direction are listed in Table~\ref{tab:Ne}. The off-nucleus density ranges roughly from 50$\sim$300\,cm$^{-3}$, but with rather large uncertainties. However, for the regions with the highest SNR ($\sim5$), E-1 and E-2, we find N$_e\approx 60$\,cm$^{-3}$.

\subsubsection{Shock Models}\label{sec:shocks}
\begin{figure*}
    \centering
    \includegraphics[width=\textwidth]{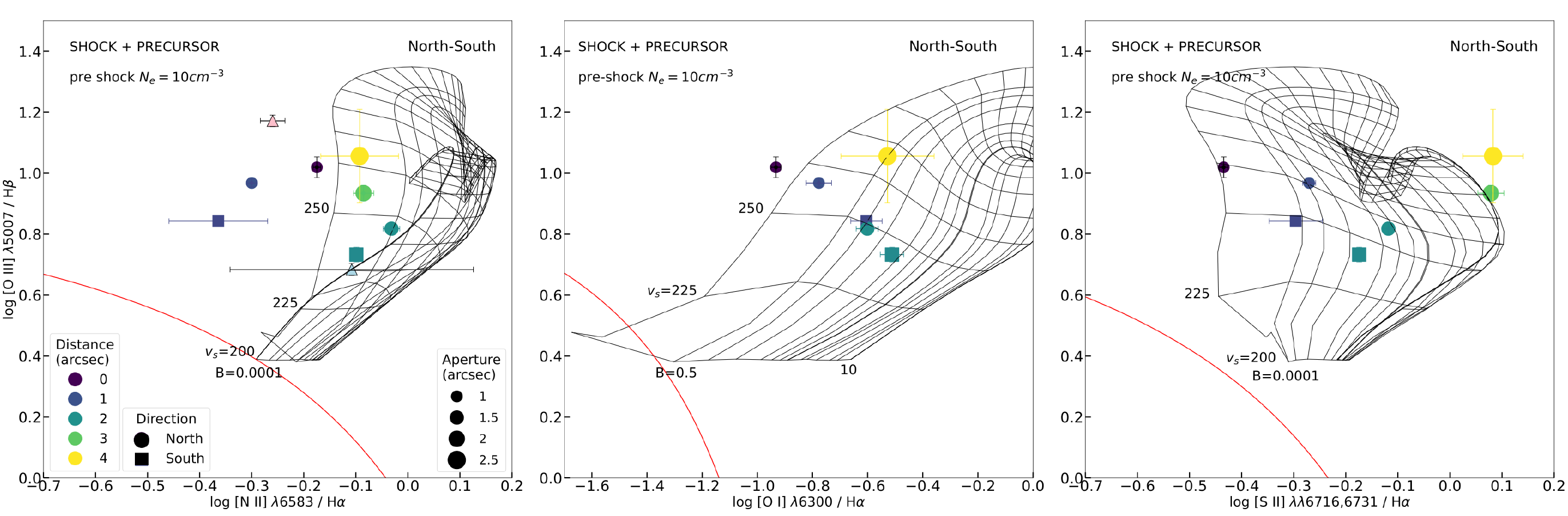}
     \caption{BPT diagrams for the nucleus and the EELR regions to the North (circles) and South (squares). From left to right, [OIII]/H$\beta$ versus [NII]/H$\alpha$, [OI]/H$\alpha$, and [SII]/H$\alpha$. The symbol color of the data points indicates distance from the nucleus and the symbol size indicates the size of the extraction aperture. A grid of radiative shock models from \citet{Allen2008} is also plotted. The model grid is annotated with the shock velocity ($v_s$) in \,km\,s$^{-1}$, which varies along the approximately vertical lines and the magnetic field strength (B), which varies along the approximately horizontal curves. We use a pre-shock density of 10\,cm$^{-3}$. The red line is the ``maximum starburst'' line of \citet{Kewley2001OpticalGalaxies}, which separates photoionization by star formation (below the line) from photoionization by AGN (above it).  The triangles represent the nucleus fit results when the blue wing component is included for the H$\alpha$, H$\beta$, and [NII] narrow lines, where, the light blue and pink symbols represent the ``wing'' (Nuc-wing)  and ``core'' (Nuc-core) components, respectively.
     }
    \label{ns_shock}
\end{figure*}

\begin{figure*}
    \centering
    \includegraphics[width=\textwidth]{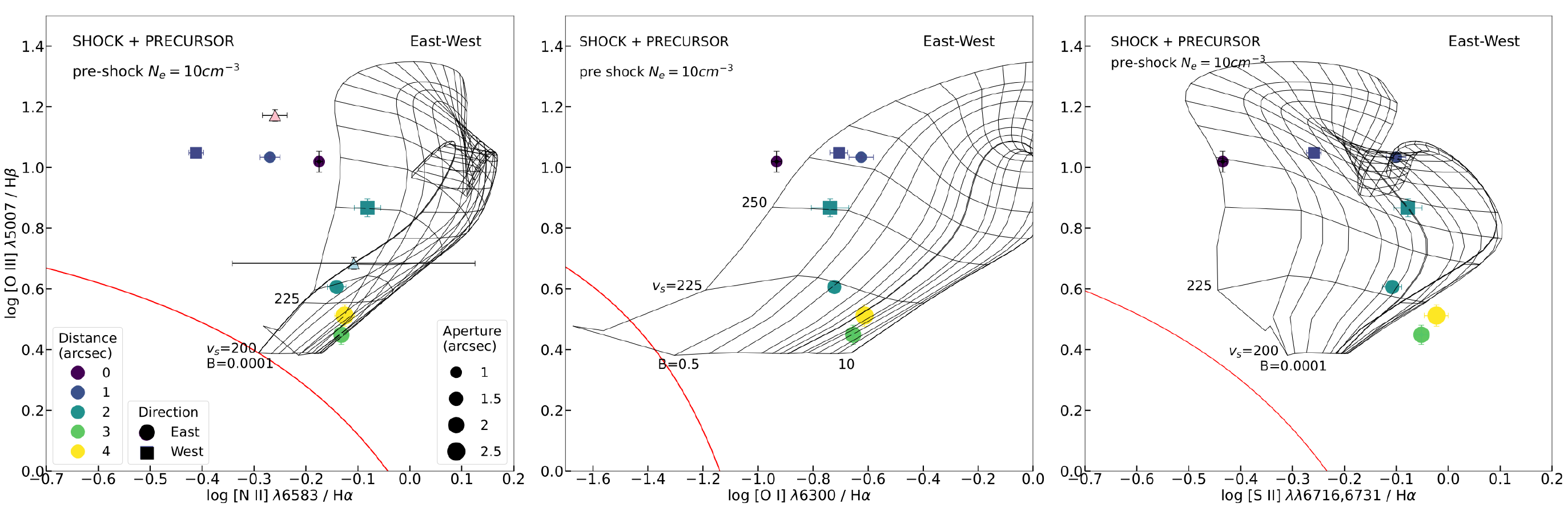}
    \caption{BPT diagrams for the nucleus and the EELR regions to the East (circles) and West (squares). From left to right, [OIII]/H$\beta$ versus, [NII]/H$\alpha$, [OI]/H$\alpha$, and [SII]/H$\alpha$. The colors and sizes of the symbols, as well as the shock model grid, are as described in the caption to Figure \ref{ns_shock}.
    }
    \label{ew_shock}
\end{figure*}

%http://3mdb.astro.unam.mx/  
To test for fast, radiative shocks, which may be caused by an outflow from the quasar, or tidal interactions with a merging galaxy, we have compared the measured line ratios with a grid of shock models in the BPT diagrams shown in Figures~\ref{ns_shock} and~\ref{ew_shock}.  The models were calculated with the \textsc{MAPPINGS V} \citep{Sutherland2017EffectsModels} photoionization and radiative shock modeling code using \citet{Allen2008}'s parameters and were obtained from the \textsc{Mexican Million Model Database} (\textsc{3Mdb}\footnote{\url{http://3mdb.astro.unam.mx/}}) \citep{Alarie2019ExtensiveDatabase,Morisset2014A3MdB}.  Here we set the abundances to Solar and the pre-shock electron density to 10\,cm$^{-3}$.  The grid covers shock velocities, $200 \leq v_s \leq 1,000$\,km\,s$^{-1}$ in increments of 50\,km\,s$^{-1}$, and magnetic field ($B$) strengths, $10^{-4}$, 0.5, 1.0, then increasing in increments of 1.0 $\mu$G to 10.0 $\mu$G.  
It was found that the pure shock models did not reproduce the measured line ratios and therefore we used models that include a photoionized precursor.  A precursor occurs for shock velocities $\gtrsim 300$\,km s$^{-1}$, when the shocked, ionized gas is hot enough ($10^{6-9}$K) to emit strongly in the EUV and X-rays. This radiation photoionizes the gas ahead of the shock, which in-turn contributes to the optical line emission. 

\begin{table}
 \caption{Weighted mean estimates of the electron density calculated from the [SII] intensity ratio from the nucleus and 1-2\arcsec extraction aperture spectra in each direction.}
\begin{tabular}{ |c|c|c| }
\hline
 Direction & N$_e$ [cm$^{-3}$] & Error [cm$^{-3}$]\\ 
 \hline
 Nucleus & 164.29 & +106.26 -141.69\\
 North & 276.53 & +182.69 -135.95 \\
 South & 59.08 & +49.32 -41.19 \\ 
 East & 56.75 & +5.10 -2.74 \\ 
 West & 193.28 & +82.09 -68.33 \\ 
 \hline
\end{tabular}
 \label{tab:Ne}
\end{table}

The majority of the data points fall within the shock+precursor grids in both the N-S (Figures~\ref{ns_shock}) and E-W directions (Figures~\ref{ew_shock}).  There is no apparent trend with distance, shock velocity, or magnetic field.  However, the data are generally consistent with $v_s=250-275$\,km\,s$^{-1}$ but no preferred magnetic field value. 

It can be argued that the redshifted kinematic component is consistent with shock ionization. In Figure \ref{ns_shock}, the locations of the N-3-lg and N-4-lg points with respect to the model grids suggest a shock velocity $\sim 250-300$\,km\,s$^{-1}$ and moderately strong magnetic fields.  To the East in Figure~\ref{ew_shock}, the E-3-lg and E-4-lg points suggest a strong magnetic field, B $\sim10\mu$G, and a lower shock velocity, $200<v_s<225$\,km\,s$^{-1}$.  Therefore overall, the redshifted gas component matches models with a range in shock velocities, $200 < v_s < 300$\,km\,s$^{-1}$, and strong magnetic fields, B $\sim10\mu$G. 

The Nuc-core component falls outside the model grid, as do Nuc-0 and all the points representing the 1\arcsec\ offset regions (N/S/E/W-1) in the [OIII]/H$\beta$ to [NII]/H$\alpha$ BPT diagram. In contrast, the blue wing (Nuc-wing) is consistent with $v_s\approx250$\,km\,s$^{-1}$, albeit with a large uncertainty in the [NII]/H$\alpha$ ratio.

The pre-shock density, N$_e = 10$\,cm$^{-3}$, adopted for the model grid shown in Figures~\ref{ns_shock} and \ref{ew_shock} seems reasonable given that the best determined EELR density  measurement is N$_e \sim 60$\,cm$^{-3}$ (Table~\ref{tab:Ne}). However, as the densities derived from the EELR spectra span a wide range, and most have large  uncertainties, we also constructed BPT diagrams for model grids with pre-shock densities 1\,cm$^{-3}$ and 100\,cm$^{-3}$. These are included in the Appendix, Figures \ref{fig:EWshock01} and \ref{fig:EWshock100}, respectively. Overall, these grids do not match distribution of the observed line ratios as well as the N$_e = 10$\,cm$^{-3}$ grid.

Assuming $B =10\mu$G and v$_s=225$\,km\,s$^{-1}$, \textsc{MAPPINGS V} predicts $F_{H\alpha}=18.8\times10^{-4}$ and $F_{H\beta}=6.4\times10^{-4}$\,erg\,cm$^{-2}$\,s$^{-1}$, per unit area of shock front.~ The 3D structure of the EELR is unknown. However for comparison with the shock model predictions, we estimated the surface brightness by assuming a 1\arcsec$\times$1\arcsec area.  From our E-2 spectrum, we estimate $F_{H\alpha}=15.1\pm0.19\times10^{-4}$ and $F_{H\beta}=4.8\pm0.14\times10^{-4}$\,erg\,cm$^{-2}$\,s$^{-1}$.~ Therefore, it is possible that shocks are energetically capable of producing the observed line fluxes in the EELR of E1821.

% NS figures

\subsubsection{Photoionization Models}
\label{sec:AGNphoto}

\begin{figure*}
    \centering
    \includegraphics[width=\textwidth]{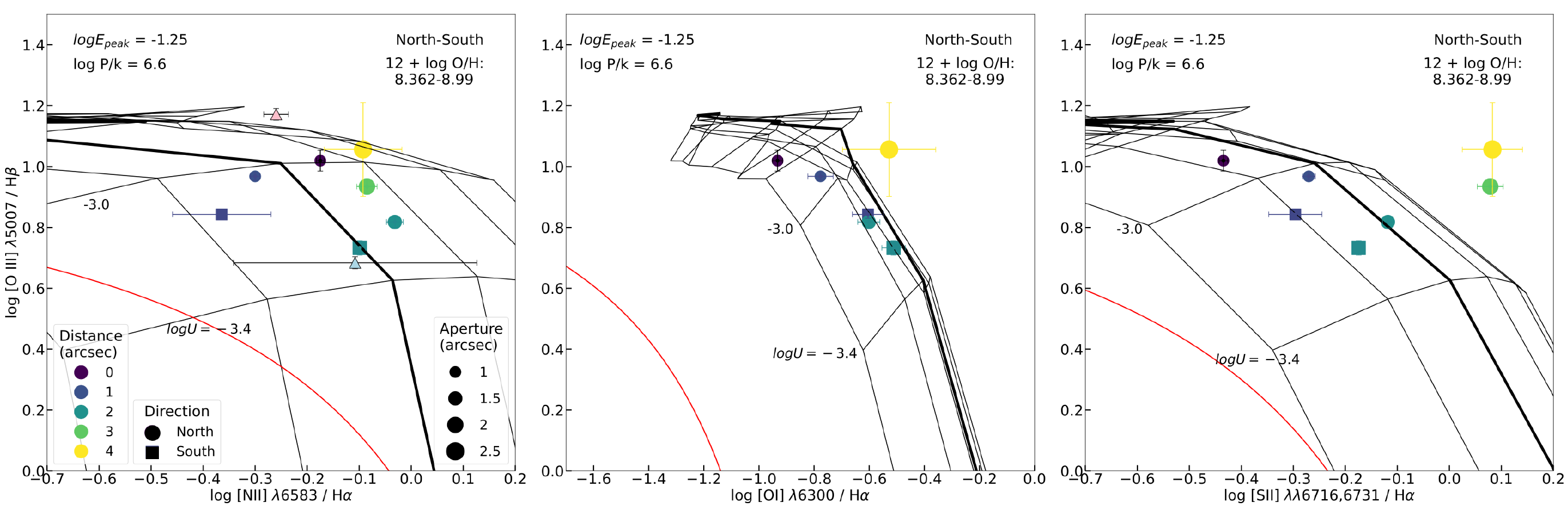}
    \caption{BPT diagrams for the nucleus and the EELR regions to the North (circles) and South (squares).  From left to right, [OIII]/H$\beta$ versus [NII]/H$\alpha$, [OI]/H$\alpha$, and [SII]/H$\alpha$. The color of the symbols indicates distance from the nucleus and the symbol size indicates the size of the extraction aperture. A grid of photoionization models from MAPPINGS V \citep{Sutherland2017EffectsModels} is also plotted using the code of \citet{Thomas2016AMODELING}.  The model grid is annotated with abundances that vary along the approximately horizontal lines with the bold line indicating solar abundance, $12+\log O/H=8.76$.   The abundances increase, left to right, from $12+\log O/H=$ 8.362 to 8.99. The log of the ionization parameter, $\log U$, varies along the approximately vertical lines. The red line is the ``maximum starburst'' line of \citet{Kewley2001OpticalGalaxies}, which separates photoionization from star formation (below the line) from photoionization by AGN (above it).  The triangles represent the nucleus fit results when the blue wing component is included for the H$\alpha$, H$\beta$, and [NII] narrow lines, where, the light blue and pink symbols represent the ``wing'' (Nuc-wing)  and ``core'' (Nuc-core) components, respectively.
    }
    \label{ns_ion}
\end{figure*}

\begin{figure*}
    \centering    
    \includegraphics[width=\textwidth]{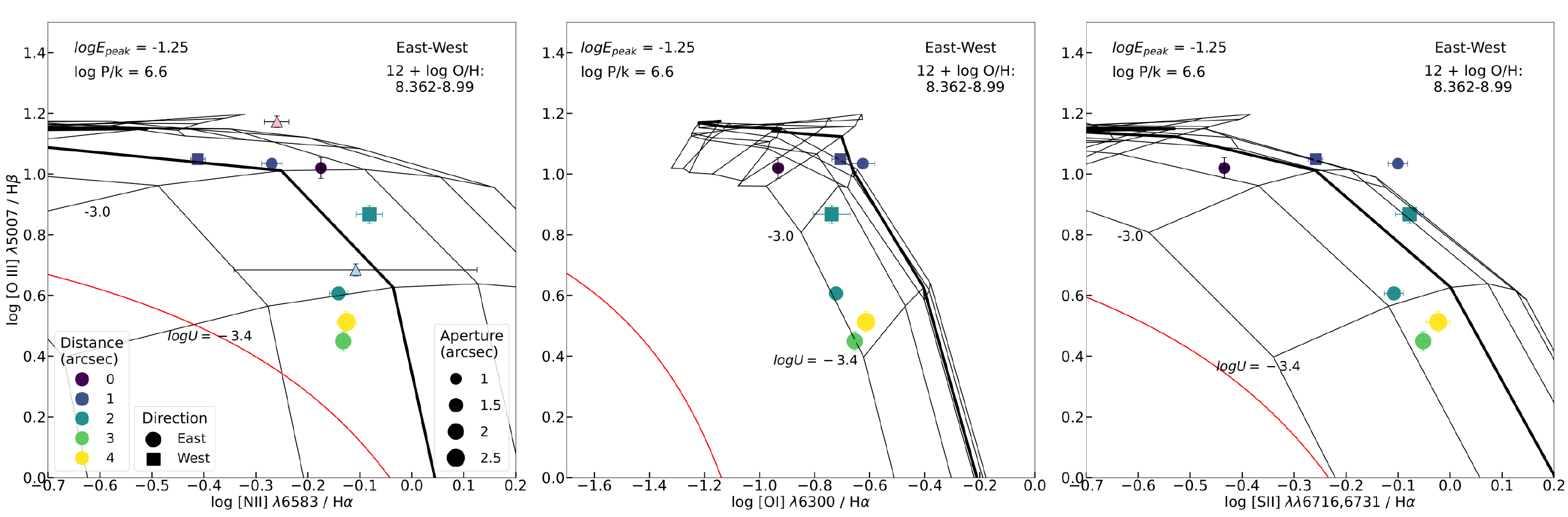}
    \caption{BPT diagram for the nucleus and the EELR regions to the East (circles) and West (squares).  From left to right, [OIII]/H$\beta$ versus [NII]/H$\alpha$, [OI]/H$\alpha$, and [SII]/H$\alpha$. The model grid, the red line, and symbols, are described in the caption to Figure \ref{ns_ion}.
    }
    \label{ew_ion}
\end{figure*}

Photoionization by the AGN ionizing continuum is likely to be the dominant ionization mechanism for the nuclear region (i.e., the AGN's NLR) and perhaps also the surrounding extended emission line regions. We have compared the measured line ratios with a grid of NLR photoionization models computed with the \textsc{MAPPINGS V} photoionization and radiative shock modeling code \citep{Alarie2019ExtensiveDatabase,Sutherland2017EffectsModels} in Figures~\ref{ns_ion} and \ref{ew_ion}.  The photoionization models are a subset of the grid supplied with the \textsc{NebulaBayes} code \citep{Thomas2018InterrogatingAbundances}. The latter code compares a set of observed line ratios with photoionization model grids, using Bayesian statistics to estimate the probability distributions of the model parameters. 

The NLR photoionization models utilize the \textsc{Oxaf} model for the AGN ionizing continuum \citep{Thomas2016AMODELING}, which includes a thermal component, representing the accretion disk and a non-thermal power-law component, attributed to inverse Compton scattering. This is described by three parameters. We follow \citet{Thomas2018InterrogatingAbundances} in fixing the proportion of flux in the non-thermal power-law component, $p_{NT}$, and the photon index of this component, $\Gamma$, to the fiducial values, $p_{NT}=0.15$ and $\Gamma = 2.0$, respectively. Lastly, the peak energy of the the accretion disk emission that best matched the data was $\log E_{peak}=-1.25$\,keV \citep{Thomas2018InterrogatingAbundances}.  We also used the \textsc{NebulaBayes} code to find the best fit values of the parameters to the nucleus data (Nuc-0), which resulted in an estimate of $\log$ E$_{peak}\sim-1.28$.  We also adopted the default value of the gas pressure, $\log$ P/k $=6.6$\,cm$^{-3}$\,K, since changing this parameter had little affect on the grid. The two free parameters of the model grid that is plotted in the BPT diagrams are the ionization parameter, $U$, and the oxygen abundance, $12+\log$ O/H, from which the other element abundances are scaled. $\log U$ ranges between -3.8 and -0.2 in intervals of 0.4. The oxygen abundance ranges from 8.362 to 8.990, with the solar value being 8.76.  

In general, the photoionization model grid encompasses the majority of the data points in all three BPT diagrams, in both the N-S and E-W directions. Although there is some scatter, the points tend to cluster along the solar abundance (bold) line, with a spread in $\log$ U of $\sim 0.4$ dex, in the range $-3.4 \lesssim \log$ U $\lesssim -3.0$. There is some evidence for a decrease in ionization parameter with distance from the nucleus to the North and South (Figures \ref{ns_ion}), with the nucleus having a higher value of the [OIII]/H$\beta$ ratio and hence higher inferred $\log U$ than the offset regions at 1 to 3\arcsec, which have the most reliable measurements. The exception is N-4-lg, which has a high [OIII]/H$\beta$ ratio $\approx 12$) but has large error bars.

Table~\ref{tab:NlogU} lists the values of $\log U$ estimated from the BPT diagrams for the nucleus and odd interval distances North of the nucleus (N-1 and N-3-lg; Figure \ref{ns_ion}) as well as corresponding values obtained with \textsc{NebulaBayes}. The values obtained from the BPT diagrams match the Bayesian estimates well, and show that $U$ tends to decrease with distance from the nucleus.  

In the E-W direction (Figure \ref{ew_ion}), there is a wider spread with $\log U$, which is clearly lower for the offset regions to the East  (E-3-lg to E-4-lg) than for the nucleus and surrounding regions.  The nucleus and adjacent regions within 2\arcsec\ lie between $\log U \sim -3.4$ and $\log U \sim -3.0$, while the offset regions further out to the East have $\log U \lesssim -3.4$. Values of $\log U$ estimated from the BPT diagrams for the nucleus and the odd interval offset regions to the East (E-1 and E-3-lg), are compared in with corresponding values obtained with \textsc{NebulaBayes} in  Table~\ref{tab:ElogU}. The Bayesian estimates and those from the BPT diagram agree well and indicate that the offset region, E-3-lg has a lower $\log U$ by $\approx 0.4$\,dex than the nucleus and E-1. 

The Nuc-core point has the highest [OIII]/H$\beta$ ratio and therefore $U$, although it is slightly off the model grid.  The Nuc-wing component has a much lower [OIII]/H$\beta$ ratio and is consistent with $\log U\sim -3.3$.

\begin{table}
 \caption{Values of $\log U$ for the spectra listed in the first column. The values obtained from \textsc{NebulaBayes}  \citet{Thomas2018InterrogatingAbundances} are compared to the approximate values estimated from Figure \ref{ns_ion}.
 }
\begin{tabular}{ |c|c|c|c|c| }
\hline
 Dir/Dis & Bayesian & $\log$ [NII]/H$\alpha$ & $\log$ [OI]/H$\alpha$ & $\log$ [SII]/H$\alpha$\\ 
 \hline
Nuc-0 & -3.05714 & -3.0 & -2.4 & -2.8\\
 N-1 & -3.05714 & -3.1 & -3.0 & -3.1\\ 
 N-3-lg & -3.34286 & -3.2 & -3.1 & -\\ 
 \hline
\end{tabular}
\label{tab:NlogU}
\end{table}

%% PHOTOIONIZATION MODELS

\begin{table}
 \caption{Values of $\log U$ for the spectra listed in the first column. The values obtained from \textsc{NebulaBayes}  \citet{Thomas2018InterrogatingAbundances} are compared to the approximate values estimated from Figure \ref{ew_ion}.
 }
\begin{tabular}{ |c|c|c|c|c| }
\hline
 Dir/Dis & Bayesian & $\log$ [NII]/H$\alpha$ & $\log$ [OI]/H$\alpha$ & log [SII]/H$\alpha$\\ 
 \hline
 Nuc-0 & -3.05714 & -3.0 & -2.4 & -2.8\\
 E-1 & -3.05714 & -3.0 & -3.0 & -3.1\\ 
 E-3-lg & -3.62857 & -3.4 & -3.4 & -3.4\\ 
 \hline
\end{tabular}
 \label{tab:ElogU}
\end{table}

\section{Discussion}
\label{sec:discussion}

We have fitted the emission lines in 1D spectra extracted at 1\arcsec~(4.5\,kpc) intervals in distance from the nucleus in the North, South, East, and West directions, to characterize the properties of the ionized gas surrounding E1821. We have detected line emission (at least in [OIII]$\lambda$5007) out to a distance $\sim 5\arcsec$~(22.5\,kpc) in all directions. We measured the velocity relative to the nucleus and the line fluxes to examine the kinematics and ionization of the EELR. We find three kinematically distinct components, which appear to be associated with different structures within the circumnuclear and extended gas distribution.

% systemic gas
One component is kinematically quiescent at approximately the systemic velocity of E1821. There is no systematic velocity shift in the regions to the S and W out to $\sim 5\arcsec$, or in the regions within $\sim 2\arcsec$ to the N and E. The velocity shifts in these regions are $\lesssim $100\,km\,s$^{-1}$, relative to the nucleus.  The core of the [OIII] line has a roughly constant FWHM $\sim 400$\,km\,s$^{-1}$ and shows no systematic variations with distance or direction. However, there is evidence for an increase in velocity dispersion, to FWHM $\sim 600$\,km\,s$^{-1}$ in regions N-2 and E-2, which border the redshifted component. Otherwise, there is no evidence for large-scale motions, or large kinematic disturbances throughout the EELR to the S and W, or within the innermost regions to the N and E. The line ratios measured in this kinematically quiescent component are consistent with AGN photoionization.

% blue wing
The [OIII]$\lambda\lambda$4959,5007 lines in the nucleus and adjacent regions within $\lesssim 2\arcsec$ have asymmetric blue wings, which we model with a separate gaussian component that is blue shifted by $\sim 200$\,km\,s$^{-1}$ relative to the line core and much broader ($\sim 1200-1400$\,km\,s$^{-1}$). As the blue wing is only confidently measured in the [OIII] lines in our spectra, we cannot determine its ionization mechanism with any certainty. 
Asymmetric blue wings are common features of the [OIII] lines in both narrow-line and broad-line AGN  \citep[e.g.,][]{Whittle1992VirialParameter,Greene2005AGALAXIES,Mullaney2013Narrow-lineConnection}. In particular, \citet{Mullaney2013Narrow-lineConnection} found that the average [OIII]$\lambda 5007$ line profile, formed by stacking $\sim 10^4$ spectra of broad-line AGN from the Sloan Digital Sky Survey (SDSS), is well fitted by two gaussian components, a narrow core with FWHM $= 335$\,km\,s$^{-1}$, and a broader component, with FWHM $=851$\,km\,s$^{-1}$, that is blueshifted by $148$\,km\,s$^{-1}$. The dispersion and blueshift of the blue wing component in E1821 are comparable with, but larger than the corresponding average values for \citeauthor{Mullaney2013Narrow-lineConnection}'s sample. This appears to be consistent with the trends for AGN with higher Eddington ratios and more radio powerful sources, such as E1821 (see below), to have broader [OIII] lines with more prominent blue wings. 

The blue asymmetric wings in [OIII] line profiles are generally considered to be indicative of outflows, which may be driven by radiation pressure, or mechanical energy input by the radio source (e.g., \citet{Alexander2012WhatHoles}). Given that E1821 is both a luminous AGN, with an Eddington ratio $\sim 0.25-0.5$ \citep{Reynolds2014THEMEDIUM}, and has an extended radio source comparable in morphology and luminosity with those of FR I radio galaxies \citep{Blundell2001TheStructure}, it is likely that blue wing is produced by a wind emanating from the nucleus. Since it is present in all four cardinal directions within the inner $\sim 2\arcsec$, we infer that the wind must have a spatial scale $\sim 10$\,kpc, a wide opening angle, and is approximately aligned with our line of sight.  We cannot establish a direct connection between the radio source and the [OIII] blue wing component; the 1\arcsec\,-scale inner jet is contained within the central 1\arcsec\, extraction aperture of the nucleus spectrum, and the slit positions are misaligned with the large-scale ($\sim 30$\arcsec) radio source. Nevertheless, the presence of this feature suggests that mechanical energy feedback is currently occurring over spatial scales comparable with effective radius of the host galaxy. In this respect, E1821 may be similar to the $z\approx 0.1$ radio-quiet type 2 quasar SDSS J165315.06+234943.0, in which there is good evidence for galaxy-wide feedback driven by the radio source \citep{Villar-Martin2017Galaxy-wideQuasar}.  

%redshifted gas
In the EELR at distances 3-5\arcsec~(13.5-22.5\,kpc) from the nucleus to the N and E, the core [OIII] component is redshifted by $\sim$ 300\,km\,s$^{-1}$ and has a FWHM of about 400\,km\,s$^{-1}$. This emission can be seen as a distinct redshifted blob in the 2D spectra (Figure \ref{2dspec}), centered at $\sim3.5\arcsec$ (15.75\,kpc) N and $\sim3.0\arcsec$ (13.5\,kpc) E of the nucleus. A comparable velocity shift is also seen in the other narrow lines.  This redshifted gas seems to be associated with a broken arc of [OIII] emission seen in the HST ramp filter image (Figure \ref{hst}) that extends from the N, $\sim2.8\arcsec$ (12.6\,kpc) from nucleus, around to the ESE, $\sim3.5\arcsec$ (15.75\,kpc) from nucleus. The locations at which the redshifts are measured in the spectra coincide with the northern and eastern extremities of this feature. 

The BPT diagrams discussed in Section~\ref{sec:AGNphoto} show that the line ratios of the redshifted component are generally consistent with AGN photoionization, with an ionization parameter $\log U\sim -3.3$. Using the measured flux of the broad H$\alpha$ line, 1.65$\times10^{-12}$erg\,s$^{-1}$\,cm$^{-2}$, and assuming a BLR covering fraction $\sim 0.1$, we use simple recombination theory to estimate the ionizing photon luminosity required to photoionize the BLR, Q$_{nuc}\sim 5.6\times$10$^{54}$\,s$^{-1}$. This can be compared with the ionizing luminosity required to photoionize the EELR. For example, using the ionization parameter derived for E-3 (Table~\ref{tab:ElogU}), at a projected distance of 13.7\,kpc from the nucleus, we find Q$_{e3}\gtrsim $3$\times$10$^{54}$\,s$^{-1}$, where we have assumed a gas density of 10\,cm$^{-3}$. Therefore, we infer that quasar's ionizing luminosity is sufficient to photoionize the EELR.  

As discussed in Section~\ref{sec:shocks}, shock models with a photoionized precursor predict line intensity ratios broadly consistent with those observed in the EELR and also appear to be energetically capable of producing the observed fluxes. Therefore, radiative shocks with velocities $\sim 200-300$\,km\,s$^{-1}$ may also contribute to the line emission.  Notably, the boundary, at E-2 and N-2, between the quiescent gas and the redshifted gas corresponds to an increase in velocity dispersion  by $\sim$200\,km\,s$^{-1}$, suggesting that strong shocks play a role in shaping the morphology of the arc-like feature.  However, examining E-2-lg and N-2-lg in the BPT diagrams, the line ratio values are not obviously different compared to other regions.  In some galaxies with large-scale outflows, the  line flux ratios vary systematically with velocity dispersion, which is consistent with ionization by radiative shocks \citep{Dopita2012The7793,Ho2014TheGalaxy,Perna2020MUSEFeedback}.  We find no clear relation between the EELR line ratios and the FWHM in E1821. However, in terms of kinematics, the emission line regions in E1821 appear to be similar to the sample of high-redshift ULIRGs hosting AGN studied by \citet{Harrison2012EnergeticActivity}. The velocity dispersions of the quiescent gas and blue wing, $FWHM\sim400$ and $1000-1400$\,km\,s$^{-1}$, respectively, are consistent with the low and high velocity dispersion [OIII] components found by \citeauthor{Harrison2012EnergeticActivity}, which they associate with galaxy or merger dynamics and outflows, respectively.  

If we accept that a blue wing component is also present for the other narrow lines, H$\alpha$, H$\beta$, and [NII], in the nucleus spectrum and take the results of the fits including these components at face value, then the core component, Nuc-core, has the highest [OIII]/H$\beta$ ratio, which implies a high ionization parameter.  This is reasonable since AGN photoionization is more likely to be the dominant ionization mechanism for gas near the nucleus. The blue wing component, Nuc-wing, has a much lower [OIII]/H$\beta$ ratio, implying a lower ionization parameter. It is also consistent with radiative shocks with velocities $\sim 250$\,km\,s$^{-1}$, within the range inferred from the EELR data described above. This suggests that the core component gas is photoionized by the AGN, whereas the blue wing may be produced by shocks in the outflow. However, with the exception of the [OIII] lines, the wing components are poorly constrained by the fits and therefore the ionization mechanism for the wind is  uncertain.

It is possible that the redshifted gas forming the arc-like structure is a remnant of a tidal tail produced by a recent or ongoing merger. \citet{Aravena2011} suggest that the CO emission that they detected $\sim 2\arcsec$ (9\,kpc) SE of the nucleus (red ellipse in Figure \ref{hst}) may be associated with a gas-rich companion galaxy that is merging with E1821, or may itself be a tidal structure from a previous interaction. They also note that the CO detection is close to a faint optical structure that is present in an HST I-band image. We do not see any evidence for extended [OIII] emission to the SE of the nucleus in the HST narrow-band image presented in this paper, and unfortunately the CO detection is located between our slit positions. However, it is possible that the arc-like structure to the NE is also a part of the interaction between the source of the CO emission and the host galaxy of E1821.  Clearly, additional deep, 2D spectroscopy that includes the SE and NE quadrants would be beneficial to further investigate the presence of tidal tails and a gas-rich companion galaxy.  

An alternative possible origin for the [OIII] arc is that it is the remnant of a shell of swept-up gas from the blowout phase of an earlier major merger, that led to a recoiling SMBH \citep{Robinson2010}. To obtain an estimate of the mass flow rate, we assume that the arc is part of an expanding spherical shell with a radius of about 3\arcsec\ ($r\approx 13.5$\,kpc). Approximating the geometry as a hemispherical shell, the mass-flow rate is then $\dot{m} = 2\pi r^2 N_H m_H v_{exp}$, where $N_{H}\approx\ N_e$ is the gas density (assuming that the gas is fully ionized Hydrogen), $m_H$ is the mass of the Hydrogen atom and $v_{exp}$, the expansion velocity. Taking the approximate median density of E-3 and N-3, $N_H\sim100$\,cm$^{-3}$, and the measured velocity values for E-3 and N-3, $v_{exp}\approx 250$\,km\,s$^{-1}$, we find $\dot{m}\approx7.2\times10^{5}$\,M$_{\odot}$\,yr$^{-1}$.  
Observations indicate that strong winds driven by starbursts can have mass flow rates of a few $\times 100$\,M$_{\odot}$\,yr$^{-1}$ \citep[e.g.,][]{Rupke2019AGalaxy}. 
\citet{Hopkins2012StellarWinds} provide an approximate relation between the total mass outflow rate, including swept-up gas, and the SFR, $\dot{m}_{SB} \sim 3(SFR/\rm{M_\odot\,yr}^{-1})^{0.7}$\,M$_\odot$\,yr$^{-1}$, which, for the SFR inferred for E1821 ($\sim 1000$\,M$_\odot$\,yr$^{-1}$; Sec.~\ref{sec:intro}), yields $\dot{m}_{SB}\approx378$\,M$_{\odot}$\,yr$^{-1}$. 
Galaxy-scale outflows driven by QSOs typically have $\dot{m}_{AGN}\lesssim 1000$\,M$_{\odot}$\,yr$^{-1}$ \citep[e.g.][]{Veilleux2013FASTHERSCHEL}, however, in some cases can be as high as $\sim 10^4$\,M$_{\odot}$\,yr$^{-1}$ \citep[e.g.,][]{Liu2013ObservationsNebulae, Leung2019The1.43.8}. 

Due to the uncertainties in geometry and density, the mass-flow rate calculated for E1821 is likely an overestimate, possibly by as much as an order of magnitude. This would still be 2-3 orders of magnitude greater than expected for a starburst, but comparable the highest mass-flow rates estimated for QSO winds.  This implies that the [OIII] arc is unlikely to be part of a starburst-driven wind shell remnant, but possibly part of an extreme QSO-driven wind. If that is the case, the inferred mass-flow rate equates to a mechanical luminosity of $1.4\times 10^{46}$\,erg\,s$^{-1}$, which is 10\% of E1821's bolometric luminosity.

The only previously published optical spectroscopy of the EELR around E1821 that we are aware of is that discussed by \citet{Fried1998}. This author presents a grism spectrum obtained from a slit of width 1.5\arcsec~ offset from the nucleus by 2.5\arcsec ($\sim$11\,kpc) N and oriented in the E-W direction. It therefore samples our extraction regions N-2 and N-3 as well as part of the [OIII] arc to the East. The [OIII]/H$\beta$ ratio in the offset spectrum is $\approx 4$, somewhat lower than the values we measure ($\sim 5-7$) from N-2 and N-3-lg. This suggests a slightly lower average ionization state over the much larger region covered by \citeauthor{Fried1998}'s spectrum\footnote{The wavelength range covered by \citet{Fried1998}'s data does not include H$\alpha$ and the other lines that we use in our BPT diagrams}. Nevertheless, \citet{Fried1998}  notes that the line ratios are consistent with photoionization by the quasar. The lines also have widths $\sim$600\,km\,s$^{-1}$, which is comparable to the velocity dispersion we measure in N-2 and N-3.  

As discussed in Section~\ref{sec:intro}, modelling of the spectral energy distribution indicates that dust heated by star formation accounts for 50\% of the enormous IR luminosity of E1821, implying an extremely high star formation rate $\sim 1000$\,M$_\odot$\,yr$^{-1}$   \citep{Farrah2002Sub-millimetreGalaxies, Aravena2011}. However, we find no evidence of a massive starburst in our optical spectra; the emission line ratios are consistent with AGN photoionization in the nucleus and throughout the extended regions sampled by our spectra. Either the star formation is heavily obscured by dust, or is occurring in a region that was not sampled by our spectra (e.g., the region of the CO detection). 

\section{Conclusions}
\label{sec:conc}

Previous work by \citet{Robinson2010} and \citet{Jadhav2021TheRecoil} shows that the broad emission lines of the quasar nucleus in E1821+643 are both kinematically and spatially offset with respect to the narrow lines; with a velocity shift $\sim 2000$\,km\,s$^{-1}$ and a spatial displacement $\sim 600$\,pc to the NW, relative to the narrow line emission. These results support the hypothesis that the SMBH in E1821 is undergoing gravitational recoil, following the coalescence of a progenitor binary, which would itself have formed as a result of a recent merger. Other work has shown that E1821 resides in a post-merger system, and is probably in the blowout phase.  

Our long-slit spectra reveal spatially resolved line emission extending to at least 20\,kpc\ from the nucleus in all 4 cardinal directions. Through spectral fitting we have identified three kinematically separate components of the narrow-line emission from the nucleus and surrounding EELR.  First, the broad, blue [OIII] wing that is confined to the inner 2\arcsec~(9\,kpc) of the nucleus we associate with a wide-angle polar wind from the quasar core.  A second component is located about 3-4\arcsec~(13-18\,kpc) from the nucleus to the North and East, and is redshifted by about 300\,km\,s$^{-1}$ with respect to the nucleus.  This feature, clearly seen in the 2D spectra and the spatial profiles, appears to be associated with the extremities of an arc-like structure in the HST [OIII] image.  Lastly, there is a kinematically quiescent component of gas to the South and West, and within $\sim 2\arcsec$ to the North and East, which shows no large scale motions, or dynamical perturbations, with the exception of an increased velocity dispersion near the boundary with the redshifted component.

The line ratios of the nuclear narrow line region (excluding the blue wing, which is only detected with certainty in [OIII]) and those of the quiescent and redshifted components of the EELR are consistent with AGN photoionization, with an ionization parameter $\log U\sim $3.0 to $-3.4$ and roughly solar, or slightly sub-solar, abundances. Fast radiative shocks with velocities $\sim 200-300$\,km\,s$^{-1}$ may also contribute to the ionization of the redshifted gas. The blue wing component is not reliably detected in lines other than [OIII], but when it is included in fits of the nucleus spectrum we find line ratios consistent with either AGN photoionization, at a low ionization parameter, or radiative shocks with velocities $\sim250$\,km\,s$^{-1}$.

The arc of ionized gas may be part of a tidal tail from a recent merger or tidal interaction. Alternatively, it may be a fragment of a shell of gas swept up by the quasar wind. The possibility of a recent merger is supported by the evidence of a recoiling SMBH. Furthermore, the CO detection reported by \citet{Aravena2011} indicates a significant amount of molecular gas centered 2\arcsec~(9\,kpc) SE of the nucleus, suggesting an ongoing merger with a gas rich companion. However, we find no evidence from the optical emission line ratios for the massive starburst implied by the very large IR luminosity of the host galaxy. 

The E1821+643 system is a highly unusual one, which may provide important insights into the growth and evolution of massive galaxies and their SMBH in cluster environments. Our long-slit spectra reveal that ionized gas, predominantly photoionized by the AGN, is widespread around the nucleus, and has kinematical and morphological features that point to a recent or ongoing merger. Unfortunately, however, our spectra do not cover the majority of the arc-like structure or the source of the CO detection. Integral field spectroscopy would be highly desirable to investigate the nature of these features and indeed, the origin of the EELR gas. More generally, E1821+643's unique and curious properties also make it an object of interest for future JWST and ALMA observations.    

\section*{Acknowledgements}
This research is based partly on observations made with the NASA/ESA Hubble Space Telescope obtained from the Space Telescope Science Institute, which is operated by the Association of Universities for Research in Astronomy, Inc., under NASA contract NAS 5-26555. These observations are associated with program G0-13385.

The work also made use of observations obtained at the inter-national Gemini Observatory, a program of NSF’s NOIRLab, which is managed by the Association of Universities for Research in Astronomy (AURA) under a cooperative agreement with the National Science Foundation. on behalf of the Gemini Observatory partnership: the National Science Foundation (United States), National Research Council (Canada), Agencia Nacional de Investigación y Desarrollo (Chile), Ministerio de Ciencia, Tecnolog\'ia e Innovaci\'on (Argentina), Minist\'erio da Ci\^encia, Tecnologia, Inova\k{c}\~oes e Comunica\k{c}\~oes (Brazil), and Korea Astronomy and Space Science Institute(Republic of Korea).

The authors also thank the reviewer for their thorough review and suggestions.

\section*{Data Availability}
The HST data can be accessed from the Mikulski Archivefor Space Telescopes (MAST) using the program ID 13385 at https://archive.stsci.edu/. The spectroscopy data can be accessed at the Gemini Observatory archive using the proposal ID GN-2010A-Q-103 at https://archive.gemini.edu/. The derived data generated in this research will be shared on reasonable request to the corresponding author.

%%%%%%%%%%%%%%%%%%%%%%%%%%%%%%%%%%%%%%%%%%%%%%%%%%



\bibliographystyle{mnras}
\bibliography{references} % if your bibtex file is called example.bib

%\begin{thebibliography}{99}
%\end{thebibliography}

%%%%%%%%%%%%%%%%%%%%%%%%%%%%%%%%%%%%%%%%%%%%%%%%%%

%%%%%%%%%%%%%%%%% APPENDICES %%%%%%%%%%%%%%%%%%%%%
\onecolumn
\appendix
\label{sec:appendix}
\section{Additional Figures}

\begin{figure}
    \centering
    \includegraphics[width=\columnwidth]{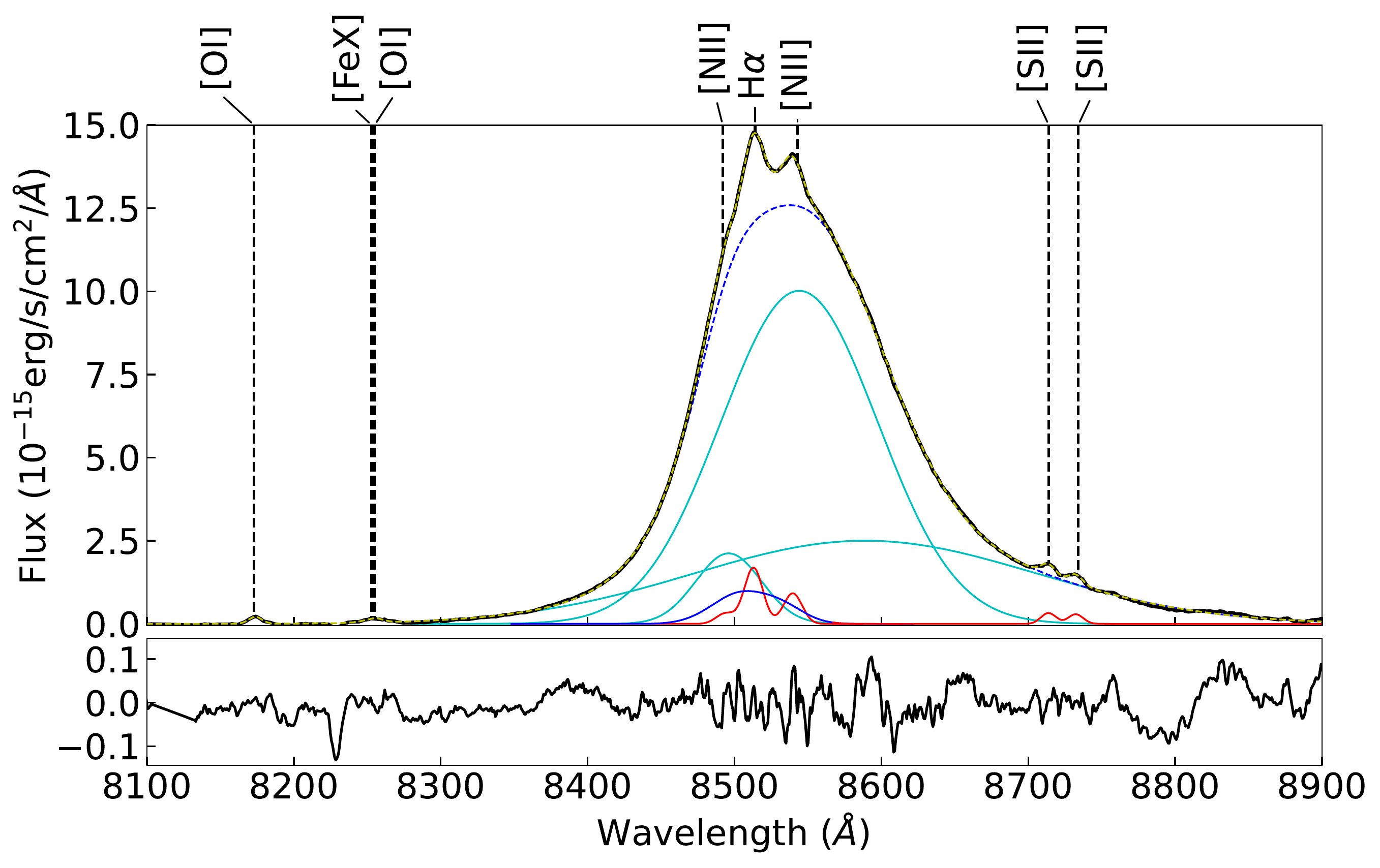}
    \caption{Same as Figure \ref{fig:Nuc_Ha}, but with the addition of a blue wing component to the  narrow H$\alpha$ and [NII] lines (blue line).  The total fit to all of the broad H$\alpha$ components is the dashed blue line.}
    \label{fig:Nuc_Hawing}
\end{figure}

\begin{figure}
    \centering
    \includegraphics[width=\columnwidth]{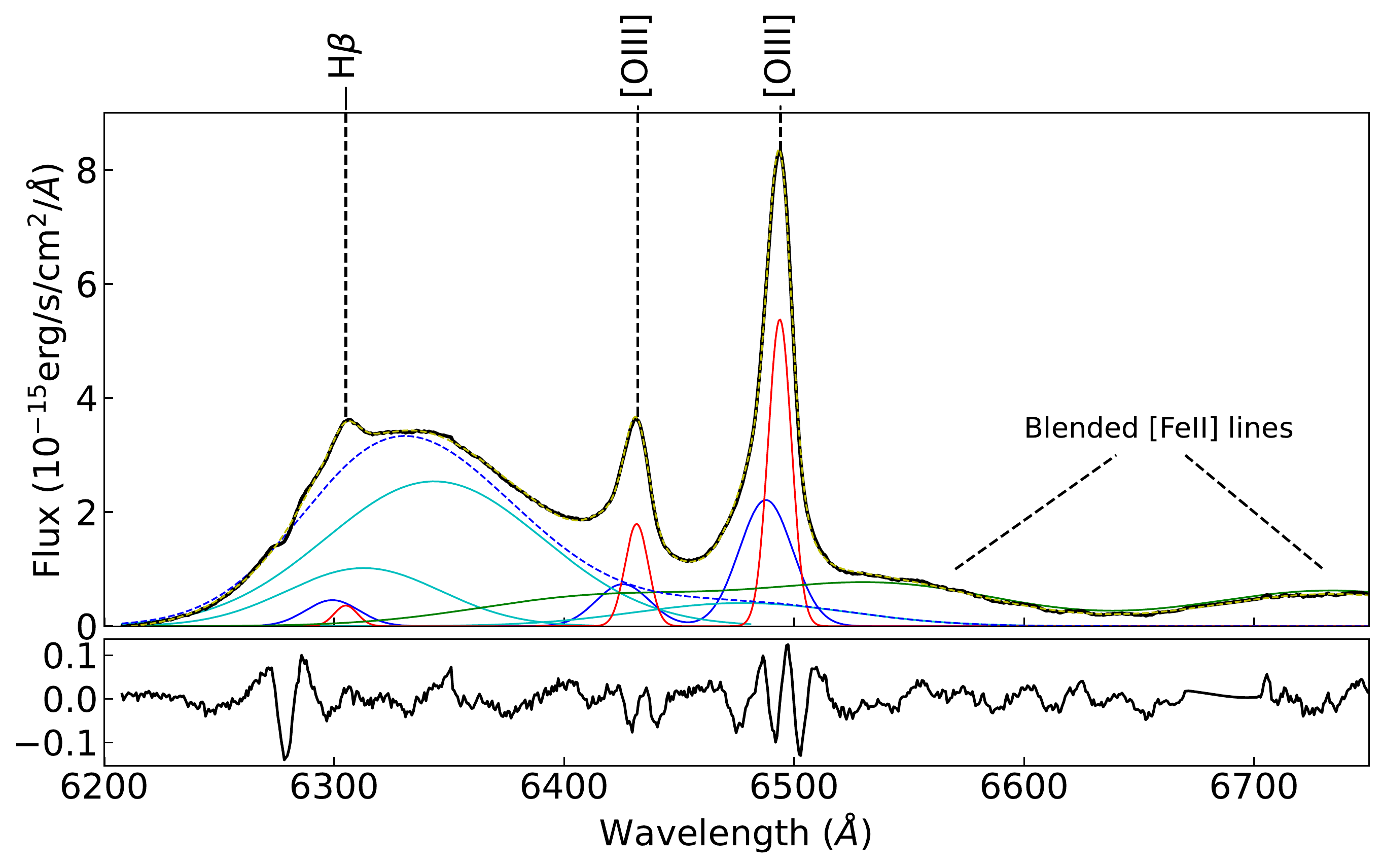}
    \caption{Same as Figure \ref{fig:Nuc_Hb}, but with the addition of a blue wing component to the narrow H$\beta$ line (blue line).  The total fit to all of the broad H$\beta$ components is the dashed blue line.} 
    \label{fig:Nuc_Hbwing}
\end{figure}

\begin{figure}
    \centering
    \includegraphics[width=\columnwidth]{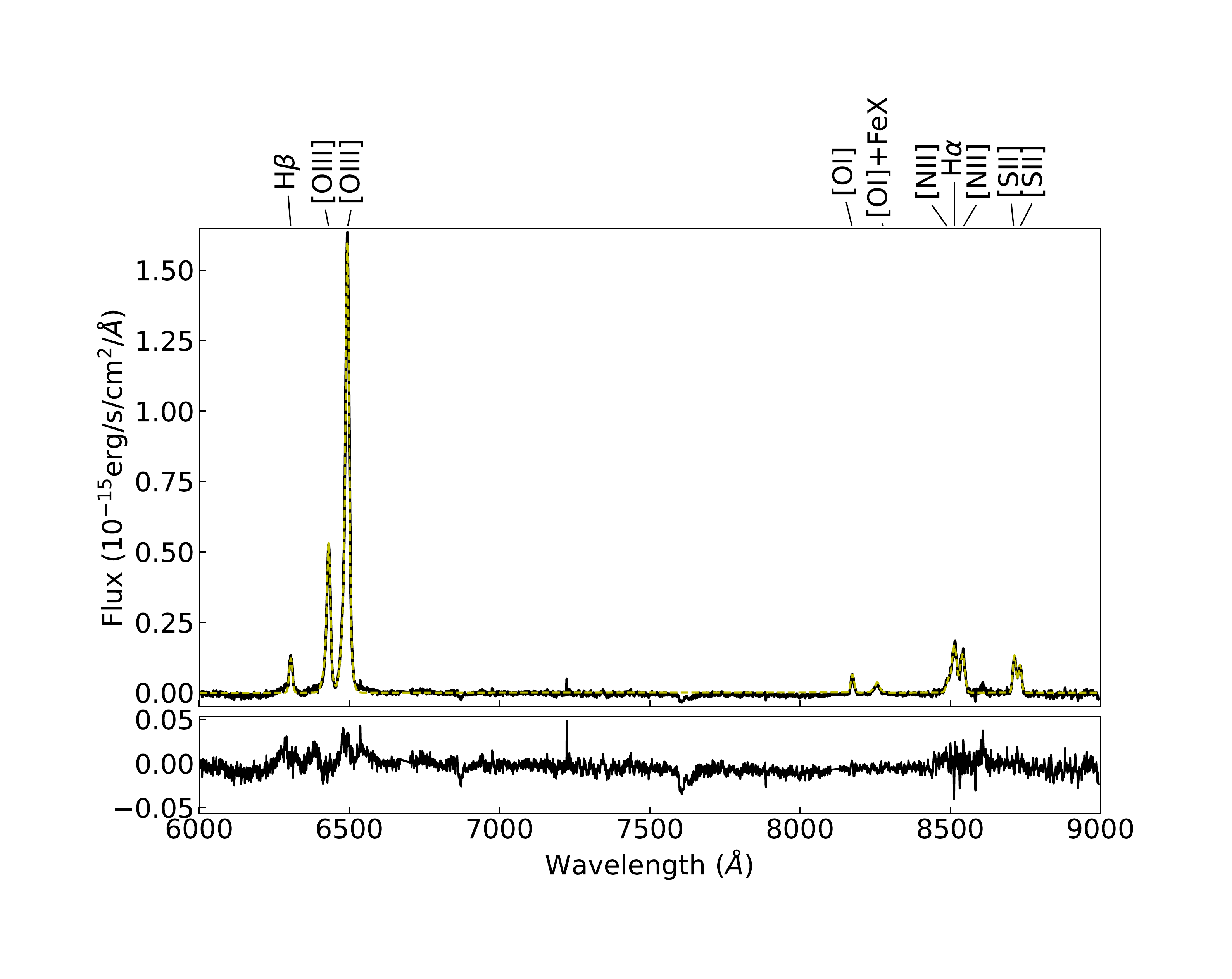}
    \caption{The complete spectrum from E-1, after subtraction of the scaled quasar PSF template spectrum (Section~\ref{sec:nucleus})}, with the narrow emission lines labeled.  The data is the black line and the overall fit is the yellow dashed line.  The lower panel shows the fit residuals.
    \label{fig:ewfull}
\end{figure}

\begin{figure}
    \centering
    \includegraphics[width=\columnwidth]{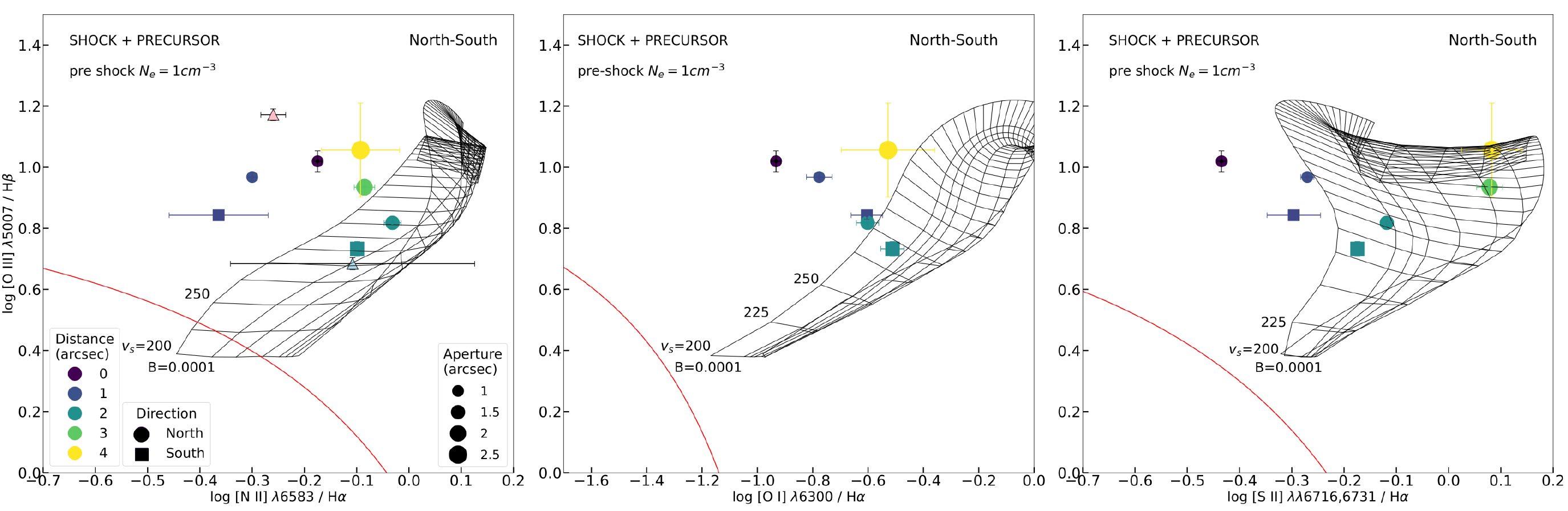}
    \caption{Same as Figure \ref{ns_shock}, but for a pre-shock density of 1\,cm$^{-3}$.}
    \label{fig:NSshock01}
\end{figure}

\begin{figure}
    \centering
    \includegraphics[width=\columnwidth]{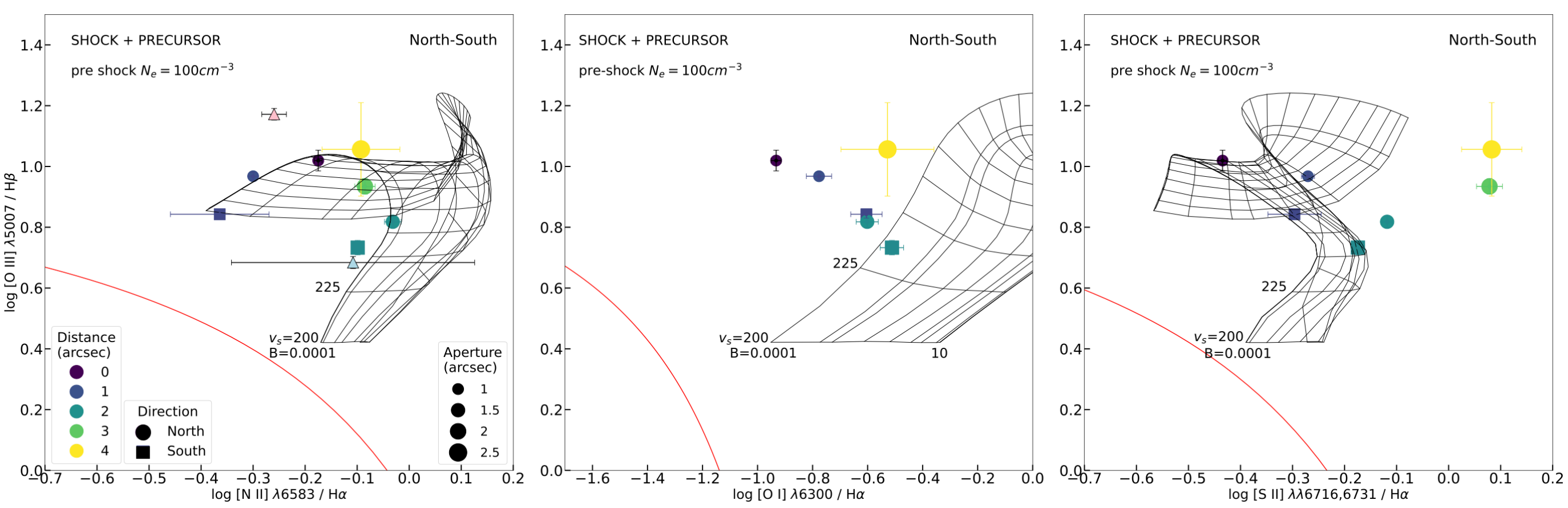}
    \caption{Same as Figure \ref{fig:NSshock01}, but for a pre-shock density of 100\,cm$^{-3}$.}
    \label{fig:NSshock100}
\end{figure}

\begin{figure}
    \centering
    \includegraphics[width=\columnwidth]{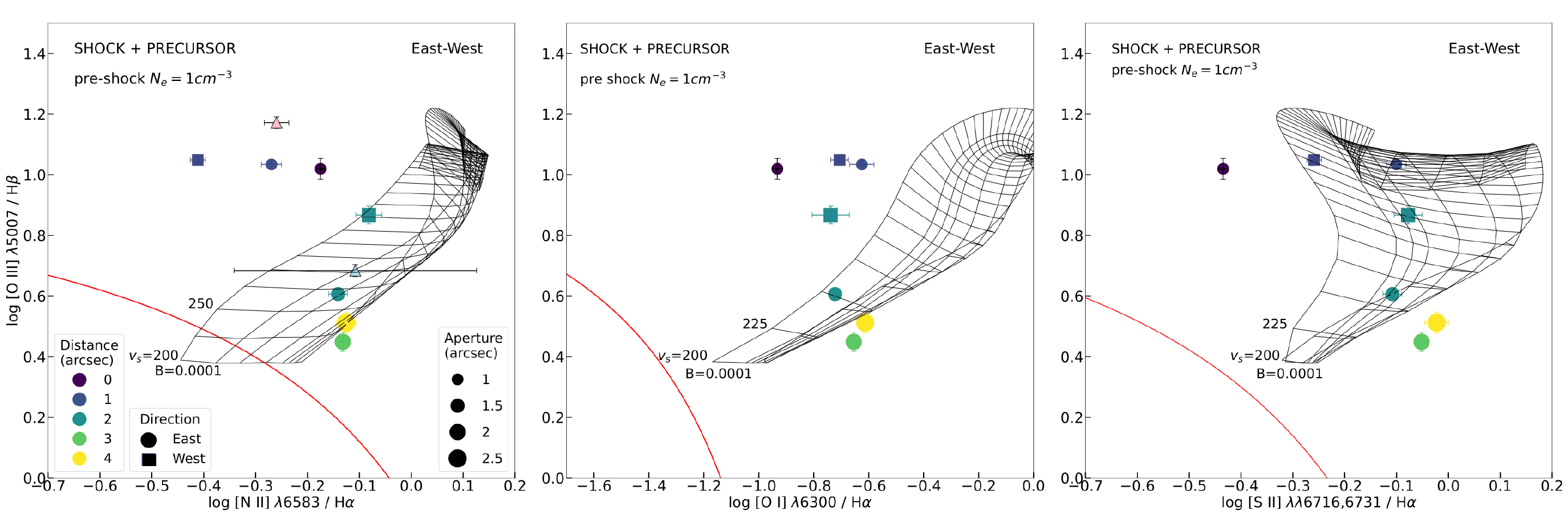}
    \caption{Same as Figure \ref{fig:NSshock01}, but for the East-West directions.}
    \label{fig:EWshock01}
\end{figure}

\begin{figure}
    \centering
    \includegraphics[width=\columnwidth]{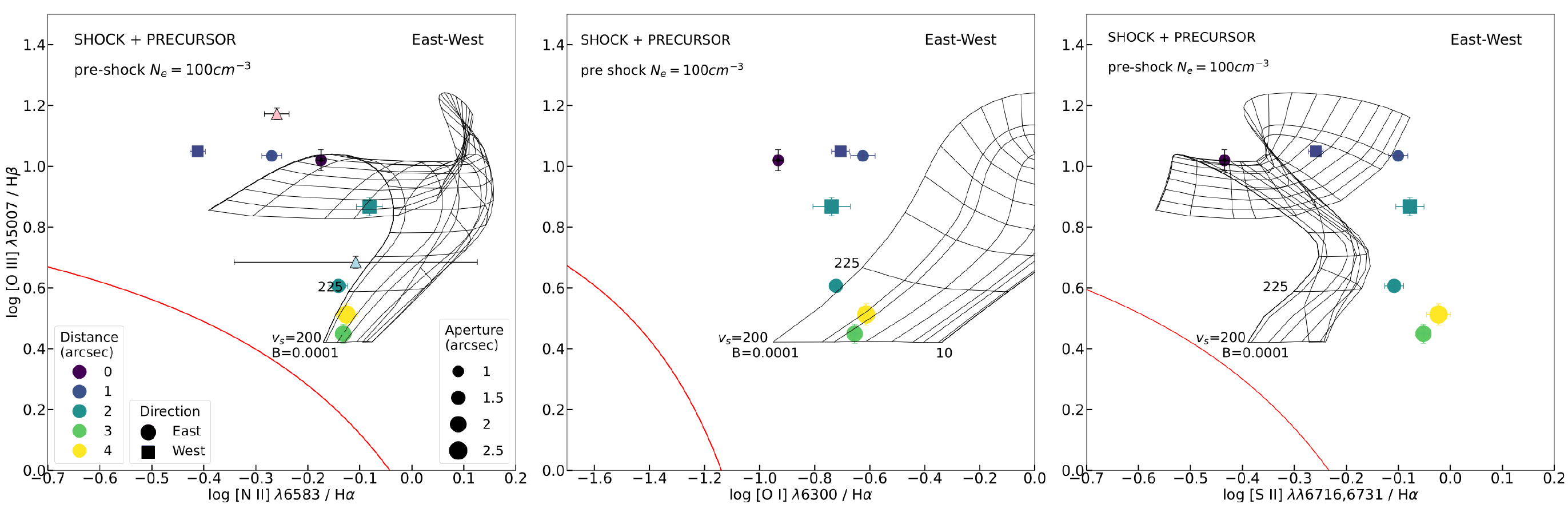}
    \caption{{Same as Figure \ref{fig:EWshock01}, but for a pre-shock density of 100\,cm$^{-3}$.}}
    \label{fig:EWshock100}
\end{figure}

%%%%%%%%%%%%%%%%%%%%%%%%%%%%%%%%%%%%%%%%%%%%%%%%%%

% Don't change these lines
\bsp	% typesetting comment
\label{lastpage}
\end{document}